\documentclass[journal]{IEEEtran}

\usepackage{ifthen}
\newboolean{DoubleColumn}
\setboolean{DoubleColumn}{true} % boolvar=true or false

\usepackage[font=footnotesize]{caption}

\usepackage{balance}
\usepackage{mathtools}
\usepackage{amsmath,amsfonts,amssymb}
\DeclareMathOperator{\sign}{sign}
\usepackage{scalerel}
\usepackage[utf8]{inputenc}
\usepackage{graphicx}
\usepackage{epstopdf}
\usepackage{ellipsis}
\usepackage{algorithmic}
\usepackage{physics}
\usepackage{svg}
\usepackage{lipsum}
\usepackage{siunitx}
\usepackage[ruled,vlined,titlenumbered,linesnumbered]{algorithm2e}
\DeclareSIUnit{\belmilliwatt}{Bm}
\DeclareSIUnit{\belisotropic}{Bi}
\DeclareSIUnit{\dBm}{\deci\belmilliwatt}
\DeclareSIUnit{\dBi}{\deci\belisotropic}
\usepackage{multirow}
\usepackage{booktabs}
\ifCLASSOPTIONcompsoc
 \usepackage[caption=false,font=normalsize,labelfont=sf,textfont=sf]{subfig}
\else
 \usepackage[caption=false,font=footnotesize]{subfig}
\fi
\usepackage[nolist]{acronym}
\usepackage[noadjust]{cite}
\usepackage{academicons}
\usepackage{tikz}
\usetikzlibrary{patterns,positioning,intersections}
\usetikzlibrary{arrows}
\usetikzlibrary{svg.path}
\usetikzlibrary{patterns}
\usetikzlibrary{calc}
\usetikzlibrary{positioning}
\usetikzlibrary{decorations.pathreplacing,decorations.markings,shapes.geometric}
\usepackage{easyReview}

\begin{acronym}
\acro{COTS}{commercial off-the-shelf}
\acro{VIDETEC}{"Erhöhte Verkehrssicherheit mittels Intelligenter Detektionstechnologien"}
\acro{BMVI}{German Ministry of Transport and Digital Infrastructure} 
\acro{mFUND}{Modernitätsfonds}
\acro{PCL}{passive coherent location}
\acro{1D}{one-dimensional}
\acro{2D}{two-dimensional}
\acro{3D}{three-dimensional}
\acro{RCS}{radar cross-section}
\acro{DFL}{device-free localization}
\acro{MDFL}{multipath-enhanced device-free localization}
\acro{FIM}{Fisher information matrix}
\acro{LoS}{line-of-sight}
\acro{GLoS}{geometrical line-of-sight}
\acro{GPS}{global positioning system}
\acro{KEST}{Kalman enhanced super resolution tracking}
\acro{EKF}{extended Kalman filter}
\acro{KF}{Kalman filter}
\acro{TWR}{two-way-ranging}
\acro{CIR}{channel impulse response}
\acro{AoA}{angle of arrival}
\acro{TDoAoMC}{time difference of arrival between multipath components}
\acro{FCC}{Federal Communications Commission}
\acro{E-911}{enhanced 911}
\acro{GNSS}{global navigation satellite system}
\acro{NLoS}{non-line-of-sight}
\acro{UWB}{ultra-wideband}
\acro{WLAN}{wireless local area network}
\acro{VT}{virtual transmitter}
\acro{VR}{virtual receiver}
\acro{VN}{virtual node}
\acro{RP}{reflection point}
\acro{RS}{reflection sequence}
\acro{MPC}{multipath component}
\acro{SBR}{single-bounce reflection}
\acro{TDOA}{time difference of arrival}
\acro{TOA}{time of arrival}
\acro{TOF}{time of flight}
\acro{RTT}{round trip time}
\acro{PDF}{probability density function}
\acro{SLAM}{simultaneous localization and mapping}
\acro{SNR}{signal-to-noise-ratio}
\acro{PF}{particle filter}
\acro{PMF}{point mass filter}
\acro{SAGE}{space-alternating generalized expectation-maximization}
\acro{MMSE}{minimum mean square error}
\acro{RMSE}{root mean square error}
\acro{MSE}{mean square error}
\acro{CDF}{cumulative distribution function}
\acro{DOP}{dilution of precision}
\acro{CRLB}{Cram\'{e}r-Rao lower bound}
\acro{PCRLB}{posterior Cram\'{e}r-Rao lower bound}
\acro{ISI}{intersymbol interference}
\acro{RBPF}{Rao-Blackwellized particle filter}
\acro{SoO}{signals of opportunity}
\acro{SISO}{single-input single-output}
\acro{SIMO}{single-input multiple-output}
\acro{RF}{radio frequency}
\acro{IMU}{inertial measurement unit}
\acro{SIS}{sequential importance sampling}
\acro{SISPF}{sequential importance sampling particle filter}
\acro{SIR}{sequential importance resampling}
\acro{SIRPF}{sequential importance resampling particle filter}
\acro{3GPP-LTE}{3rd generation partnership project - long-term evolution}
\acro{TDoAbMC}{time difference of arrival between multipath components}
\acro{subPF}{subordinate particle filter}
\acro{superPF}{superordinate particle filter}
\acro{DLL}{delay locked loop}
\acro{RFID}{radio  frequency  identification}
\acro{RSS}{received signal strength}
\acro{AWGN}{additive white Gaussian noise}
\acro{MUSIC}{multiple signal classification}
\acro{EM}{expectation maximization}
\acro{ML}{maximum likelihood}
\acro{PSD}{power spectral density}
\acro{SRA}{super resolution algorithm}
\acro{MAP}{maximum a posteriori}
\acro{LS}{least square}
\acro{DBN}{dynamic Bayesian network}
\acro{MEMS}{micro-electro-mechanical system}
\acro{INS}{inertial navigation systems}
\acro{OFDM}{orthogonal frequency-division multiplexing}
\acro{HMM}{hidden Markov model}
\acro{ZUPT}{zero velocity updates}
\acro{DMC}{Dense Multipath Components}
\acro{TS}{track-section}
\acro{log-domain}{logarithmic domain}
\acro{lin-domain}{linear domain}
\acro{bframe}[b-frame]{body frame}
\acro{iframe}[i-frame]{ineratial frame}
\acro{nframe}[n-frame]{navigation frame}
\acro{logCDF}[log-CDF]{cumulative distribution function in \ac{log-domain}}
\acro{logweight}[log-weight]{log-domain weight}
\acro{Lin-PF}[Lin-PF]{linear domain PF}
\acro{Log-PF}[Log-PF]{logarithmic domain PF}
\acro{LogMAP}[Log-PF MAP]{\ac{MAP} of Log-PF}
\acro{LogMSE}[Log-PF \acs{MMSE}]{\ac{MMSE} of Log-PF}
\acro{LogMAX}[LogPF MAX]{\ac{MAX} of LogPF}
\acro{LinMAP}[PF MAP]{\ac{MAP} of linear domain \ac{PF}}
\acro{LinMSE}[PF \acs{MMSE}]{\ac{MMSE} of linear domain \ac{PF}}
\acro{LinMAX}[PF MAX]{\ac{MAX} of linear domain \ac{PF}}
\acro{Receiver1}[receiver 1]{receiver 1}
\acro{Receiver2}[receiver 2]{receiver 2}
\acro{Receiver3}[receiver 3]{receiver 3}
\acro{MovIMU}[IMU-Transition-Model]{IMU-Transition-Model}
\acro{MovNoIMU}[Gaussian-Transition-Model]{Gaussian-Transition-Model}
\acro{algoChannelSLAMNoIMU}[NoIMU-Channel-SLAM]{NoIMU-Channel-SLAM}
\acro{RedChannelSLAM}[Dynamic-Channel-SLAM]{Dynamic-Channel-SLAM}
\acro{RBPFChannelSLAM}[RBPF-Channel-SLAM]{RBPF-Channel-SLAM}
\acro{LPF}{likelihood particle filter}
\end{acronym} 
\acresetall % reset acronyms after abstract!

\definecolor{orcidlogocol}{HTML}{A6CE39}
\tikzset{
  orcidlogo/.pic={
    \fill[orcidlogocol] svg{M256,128c0,70.7-57.3,128-128,128C57.3,256,0,198.7,0,128C0,57.3,57.3,0,128,0C198.7,0,256,57.3,256,128z};
    \fill[white] svg{M86.3,186.2H70.9V79.1h15.4v48.4V186.2z}
                 svg{M108.9,79.1h41.6c39.6,0,57,28.3,57,53.6c0,27.5-21.5,53.6-56.8,53.6h-41.8V79.1z M124.3,172.4h24.5c34.9,0,42.9-26.5,42.9-39.7c0-21.5-13.7-39.7-43.7-39.7h-23.7V172.4z}
                 svg{M88.7,56.8c0,5.5-4.5,10.1-10.1,10.1c-5.6,0-10.1-4.6-10.1-10.1c0-5.6,4.5-10.1,10.1-10.1C84.2,46.7,88.7,51.3,88.7,56.8z};
  }

}

\newcommand\orcidicon[1]{\href{https://orcid.org/#1}{~\mbox{\scalerel*{
\begin{tikzpicture}[yscale=-1,transform shape]
\pic{orcidlogo};
\end{tikzpicture}
}{|}}}}

\usepackage[hidelinks]{hyperref}

\definecolor{gray1own}{rgb}{0.7,0.7,0.7}%
\DeclareRobustCommand\mytikzLineDotted{\tikz[baseline=-0.5ex]{\draw[dotted, thick] (0,0) -- (0.5cm,0);}}
\DeclareRobustCommand\mytikzLineSolid{\tikz[baseline=-0.5ex]{\draw[] (0,0) -- (0.5cm,0);}}
\DeclareRobustCommand\mytikzLineDashGray{\tikz[baseline=-0.5ex]{\draw[color=gray1own,very thick,densely dashed] (0,0) -- (0.5cm,0);}}
\DeclareRobustCommand\mytikzLineSolidThick{\tikz[baseline=-0.5ex]{\draw[very thick] (0,0) -- (0.5cm,0);}}
\DeclareRobustCommand\mytikzNODE{\tikz[baseline=-0.5ex]{\draw[black,fill=black] (0,0) circle (.5ex);}}
\DeclareRobustCommand\mytikzONE{\tikz[baseline]{\node[shape=circle,draw,inner sep=2pt,anchor=base,scale=0.65] {1};}}
\DeclareRobustCommand\mytikzTWO{\tikz[baseline]{\node[shape=circle,draw,inner sep=2pt,anchor=base,scale=0.65] {2};}}
\DeclareRobustCommand\mytikzTHREE{\tikz[baseline]{\node[shape=circle,draw,inner sep=2pt,anchor=base,scale=0.65] {3};}}
\DeclareRobustCommand\mytikzFOUR{\tikz[baseline]{\node[shape=circle,draw,inner sep=2pt,anchor=base,scale=0.65] {4};}}

\title{Empirical Fading Model and Bayesian Calibration for Multipath-Enhanced Device-Free Localization}
\author{
\IEEEauthorblockN{Martin~Schmidhammer\textsuperscript{\orcidicon{0000-0002-9345-142X}},~\IEEEmembership{Member,~IEEE},
Christian~Gentner\textsuperscript{\orcidicon{0000-0003-4298-8195}},
Michael~Walter\textsuperscript{\orcidicon{0000-0001-5659-8716}},~\IEEEmembership{Senior~Member,~IEEE},
Stephan~Sand\textsuperscript{\orcidicon{0000-0001-9502-5654}},~\IEEEmembership{Senior~Member,~IEEE},
Benjamin~Siebler\textsuperscript{\orcidicon{0000-0002-1745-408X}},~\IEEEmembership{Member,~IEEE},
Uwe-Carsten~Fiebig\textsuperscript{\orcidicon{0000-0003-2736-1140}},~\IEEEmembership{Senior~Member,~IEEE}\\}

\thanks{The authors are with the Institute of Communications and Navigation, German Aerospace Center (DLR), 82234 Wessling, Germany (e-mail: \mbox{martin.schmidhammer@dlr.de;} christian.gentner@dlr.de; m.walter@dlr.de; stephan.sand@dlr.de; benjamin.siebler@dlr.de; uwe.fiebig@dlr.de).}
}

\begin{document}

\maketitle
\IEEEoverridecommandlockouts
\IEEEpubid{\scriptsize This work is planned to be submitted to the IEEE for possible publication. Copyright may be transferred without notice, after which this version may no longer be accessible.}

\begin{abstract}
%\Ac{MDFL} systems infer presence and location of mobile users based on the induced changes in the received power of both \aclp{MPC} and \acl{LoS} signal components.
%Thus, 
The performance of \acl{MDFL} severely depends on the information about the propagation paths within the network.
While known for the \acl{LoS}, the propagation paths have yet to be determined for \aclp{MPC}.
This work provides a novel Bayesian calibration approach for determining the propagation paths by estimating \aclp{RP}.
Therefore, first a statistical fading model is presented, that describes user-induced changes in the received signal of \aclp{MPC}.
The model is derived and validated empirically using an extensive set of wideband and \acl{UWB} measurement data.
% Second, the Bayesian approach is presented, which relates measured changes in the power of a \acl{MPC} to the location of the \acl{RP}.
Second, the Bayesian approach is presented, which, based on the derived empirical fading model, relates measured changes in the power of a \acl{MPC} to the location of the \acl{RP}.
Exploiting the geometric properties of \acl{MPC}s caused by \aclp{SBR}, the solution space of possible locations of \aclp{RP} is constrained to the delay ellipse.
%locations of the corresponding delay ellipse.
Thus, a one-dimensional elliptic estimation problem can be formulated, which is solved using a \acl{PMF}.
% This allows the formulation of a one-dimensional elliptic estimation problem, which is solved using a \acl{PMF}.
%by approximating the posterior density by a deterministic grid.
The applicability of the proposed approach is demonstrated and evaluated based on measurement data.
Independent of the underlying measurement system, the Bayesian calibration approach is shown to robustly estimate the locations of the \aclp{RP} in different environments.
\end{abstract}


\acresetall

\section{Introduction}
Our homes~\cite{viani2013}, our cities~\cite{jiong2014}, and our industries~\cite{wollschlaeger2017} are becoming smart.
%Our homes, our cities, our industries~\cite{viani2013,jiong2014,wollschlaeger2017}.
Primarily, this trend is driven by the increasing, nigh ubiquitous connectivity.
The number of connected devices is steadily growing, and with it the demand for location-based services~\cite{shit2019}.
This demand in location-awareness can be served for users carrying a localization device, e.g., by active \ac{RF}-based localization systems~\cite{sand2014}.
However, users are not always equipped with such a device, but the demand in location-awareness still exists.
Possible use cases range from monitoring applications for health and elderly care, to home security and intruder detection, to audio applications for HiFi~\cite{Kasher2020bf}.
% and use cases range from 
% intruder detection, 
% elderly care and assiseted
% smart climate control
% Hi-Fi functionalities
%applications for integrated sensing and communications. 
% Therefore, alternative passive localization systems are required, that provide location-awareness also for passive users that are not equipped with active localization devices.
Therefore, alternative passive localization systems are required enabling the localization of passive users that are not equipped with active localization devices.
%Addressing the demand for location-awareness for passive users that are not equipped with active localization devices requires alternative, passive localization systems.
\IEEEpubidadjcol
As wireless connectivity increases, \ac{RF}-based passive localization or sensing systems are becoming a viable option.
Here, the presence and location of users can be estimated by means of \ac{RF} sensor networks, exploiting the physical impact of the user on \ac{RF}-signals~\cite{patwari2010}. 
We basically distinguish between radar systems using the properties of signals directly scattered off the user, e.g.,~\cite{griffiths2005}, and \ac{DFL} systems using power changes of received signals within a wireless network caused by user-induced diffraction and shadowing effects, e.g.,~\cite{wilson2010}.
%Thereby, typical \ac{DFL} systems are realized by measuring the \ac{RSS} between the network nodes.
For inferring presence and location of users, current narrowband \ac{DFL} systems typically measure the \ac{RSS} between the network nodes in \ac{LoS}~\cite{patwari2010,wilson2010,guo2015,wang2015,kaltiokallio2017_arti,hillyard2020,kaltiokallio2021}.
The location of the user can then be either estimated by computing propagation field images, which is also known as radio tomographic imaging~\cite{wilson2010}, or by modeling the changes in the received power as a function of the user location.
The location estimation problem of the latter is then typically solved using Bayesian filtering, as in~\cite{kaltiokallio2021}.
\IEEEpubidadjcol

In the past, \ac{RSS}-based \ac{DFL} systems have attracted considerable attention since narrowband measurements are obtained from many commercially available wireless devices~\cite{wang2015}.
%, including Wi-Fi access points.
%\ac{RSS}-based \ac{DFL} systems have attracted great attention, since the measuremnts can be obtained from most commodity wireless devices including Wi-Fi access points and other low-cost transceivers.
However, recent advances in wireless communications show the trend towards wideband and \ac{UWB} technologies, which are already on the market. 
%However, recent trends in wireless communications show the advance of wideband and \ac{UWB} technologies already on the market. 
% exploiting MPC
For signals of higher bandwidth capacities, individual \acp{MPC} caused by scattering and reflection from the surrounding environment can be resolved and the corresponding signal parameters can be estimated.
The work in~\cite{beck2016} provides an example for \ac{DFL} using \ac{UWB} devices.
Here, the wide signal bandwidth is used to isolate the \ac{LoS} signal for mitigating distortions due to multipath propagation.
% For signals of higher bandwidth capacities, individual \acp{MPC} caused by scattering and reflection from the surrounding environment can be resolved and the corresponding signal parameters.
In~\cite{schmidhammer2020}, we have shown that users also induce fading to the received power of \acp{MPC}.
Based on these findings, we proposed a novel \ac{MDFL} approach in~\cite{schmidhammer2021_awpl}, since the propagation paths of \acp{MPC} inherently differ from those in \ac{LoS}.
% Since users also induce fading to the received power of \acp{MPC}, we have proposed a novel \ac{MDFL} approach in~\cite{schmidhammer2021_awpl}. 
We demonstrated that the localization performance improves without additional infrastructure by exploiting \acp{MPC} as complementary \ac{DFL} network links.
%We could show that considering propagation paths of \acp{MPC} as complementary \ac{DFL} network links improves localization performance without additional infrastructure.
%That is, considering propagation paths of \acp{MPC} as complementary \ac{DFL} network links has been shown to improve localization performance.
In our previous work, we assume prior information about the surrounding environment, e.g., provided by a floor plan, for determining the propagation paths during an initial calibration.
%In, we assume prior information about the surrounding environment, e.g., provided by a floor plan, for calculating virtual nodes to determine the propagation path.
In this work, we address the problem of determining the propagation paths, even when no prior information about the surrounding environment is available or has altered after calibration.
We thus describe a novel Bayesian calibration approach, that estimates the \acp{RP} of \ac{MPC} caused by \aclp{SBR}.

%For the proposed calibration approach, we measure the change in the received power of an \ac{MPC}, which needs to be related to the current location of the user.
For estimating the \acp{RP}, the proposed Bayesian calibration approach relates measured changes in the received power of an \ac{MPC} to the current location of the user in the network.
%For estimating a reflection point, we measure the change in the received power of an \ac{MPC}, which is then related to the current location of the user.
Note that this relation is similarly required for model-based \ac{DFL} and \ac{MDFL} approaches.
That means, we can apply similar fading models that express the power changes as a function of the user location.
For the \ac{RSS}, i.e., the received power of the direct link between transmitting and receiving node, there is a large number of fading models.
We can thereby distinguish between theoretical propagation models based on diffraction theory, e.g.,~\cite{wang2015,rampa2015,rampa2017}, and empirical propagation models, that statistically model perturbations of the received power, e.g.,~\cite{guo2015,wilson2012,hillyard2020}.
Regarding the received power of \acp{MPC}, only few models exist.
In~\cite{schmidhammer2020}, we have adapted the diffraction based model of~\cite{rampa2015} for multipath propagation.
Requiring a computational more efficient model, in~\cite{schmidhammer2021_awpl} and \cite{schmidhammer2021}, we have approximated the fading of \acp{MPC} by an exponential model.
In this work, we analyze, adapt, and evaluate the exponential model empirically using an extensive set of wideband and \ac{UWB} measurement data.

% high bandwidth allows the estimation of multipath components and the corresponding signal parameters.
% channel state information contain respective information.
% multipath-enhanced device free localization as application which can be applied using standardized signals (11bf)

%\subsection{Main Contribution}
The main contributions of this work are
\begin{itemize}
    \item the empirical derivation and validation of a statistical fading model for \aclp{MPC},
    \item the novel Bayesian calibration approach for \ac{MDFL} that estimates \aclp{RP} without a priori information about the surrounding environment,
    \item the demonstration and evaluation of the proposed Bayesian calibration approach using measurement data. 
\end{itemize}
% \begin{itemize}
% \item extensive measurement based validation of the empirical exponential model for mulitpath components (!)
% \item Bayesian estimation/mapping of reflection points as calibration of MDFL
% \item measurement based evaluation and demonstration of the calibration method
% %\item simulation based evaluation of MDFL calibrated with the novel calibration method
% \end{itemize}

The remainder of this paper is organized as follows. 
Required preliminaries including network and propagation model are given in Section~\ref{sec:pre}.
The empirical fading model is derived and evaluated in Section~\ref{sec:model}.
The novel Bayesian calibration approach is described in Section~\ref{sec:bayes_mapping} and evaluated using measurement data in Section~\ref{sec:eval}.
Finally, Section~\ref{sec:conclusion} concludes the paper.
\section{Preliminaries}
\label{sec:pre}
\subsection{Network and Propagation Model}
\label{sec:preNet}
For the \ac{MDFL} system, we consider a network of $N_{\mathrm{Tx}}$ transmitting and $N_{\mathrm{Rx}}$ receiving nodes. 
Transmitting and receiving nodes can be collocated or individually placed at known locations 
${\boldsymbol{r}_{\mathrm{Tx}_i}}$, ${i\in \{1,\dots,N_{\mathrm{Tx}}\}}$, and
${\boldsymbol{r}_{\mathrm{Rx}_j}}$, ${j\in \{1,\dots,N_{\mathrm{Rx}}\}}$, respectively.
The index set $\mathcal{P}$ determines the link configuration of the network, with link ${(i,j) \in \mathcal{P}}$ being composed of the \mbox{$i$-th} transmitting and the \mbox{$j$-th} receiving node and indexed by $l\in\{1,\dots,\left|\mathcal{P}\right|\}$.
Further, we model the received signal of link~$l$ as a superposition of scaled and delayed replica of a known transmit signal $s_l(t)$~\cite{molisch2009}.
Due to reflections and scattering of the surrounding environment, these comprise the \ac{LoS} component and a finite number of $N_l$ \acp{MPC}.
Thus, we express the received signal as
\begin{equation}
\label{eq:signal_model}
y_{l}(t) = \sum_{n=0}^{N_{l}}\, \alpha_{l,n}(t) \, s_l(t-\tau_{l,n}) + n_{l}(t),
\end{equation}
with $\alpha_{l,n}(t)$ as time-variant, complex amplitude, $\tau_{l,n}$ as static propagation delay of the $n$-th \ac{MPC}, where $n=0$ refers to the \ac{LoS} component, and $n_{l}(t)$ as white circular symmetric normal distributed noise with variance $\sigma^2_{y_l}$.
The product of the propagation delay $\tau_{l,n}$ and the speed of light $c$ defines the geometric length of the propagation path as
%The propagation delay $\tau_{l,n}$ further defines the geometric length of the propagation path as
\begin{equation}
\label{eq:prop_dist}
d_{l,n} = c \, \tau_{l,n},
\end{equation}
where for the \ac{LoS} the length of the propagation path equals the distance between the transmitting and receiving node
\begin{equation}
\label{eq:prop_dist_LOS}
d_{l} = d_{l,0} = c \, \tau_{l,0} = \lVert\boldsymbol{r}_{\mathrm{Rx}_{j}}-\boldsymbol{r}_{\mathrm{Tx}_{i}}\rVert.
\end{equation}

\subsection{Virtual Nodes and Reflection Points}
\label{sec:preVN}
For localization, the \ac{MDFL} system requires information about the geometric propagation paths between the individual network nodes.
While for \ac{LoS} components the propagation paths are inherently defined by the locations of the respective transmitting and receiving nodes, the propagation paths of \acp{MPC} still have to be determined.
In \cite{schmidhammer2021_awpl}, we have used virtual nodes, i.e., \acp{VT} and \acp{VR}, for modeling the propagation paths.
Thereby, the virtual nodes are constructed by successive mirroring of the physical nodes on reflecting surfaces.
Corresponding \acp{VT} and \acp{VR} form equidistant pairs of nodes, whose distances equal the length of the propagation delay~\cite{meissner2014,gentner2016,schmidhammer2021_awpl}.
The paths between these corresponding pairs of nodes intersect at the physical \acp{RP} on the reflecting surfaces, as exemplary shown in Fig.~\ref{fig:vt_vr_sbr}.
Thus, similar to optical ray-tracing, we can geometrically reconstruct the physical propagation paths~\cite{meissner2014}.

%The locations of corresponding \acp{VT} and \acp{VR} are denoted by
%$\boldsymbol{r}_{\mathrm{VT}_{l,n}}^{(u)}$~and~$\boldsymbol{r}_{\mathrm{VR}_{l,n}}^{(N_{l,n}-u)}$,
%where $N_{l,n}$ defines the reflection order and 
%$u\in \{0,\dots,N_{l,n}\}$ 
%indexes equidistant pairs of nodes.
%Note that here the physical transmitting and receiving nodes are referred to as
%$\boldsymbol{r}_{\mathrm{VT}_{l,n}}^{(0)}$ and 
%$\boldsymbol{r}_{\mathrm{VR}_{l,n}}^{(0)}$.

\subsection{Geometric Properties for Single-Bounce Reflections}
\label{sec:preGeo}
\begin{figure}
    \centering
   	\includegraphics[width=3.49in]{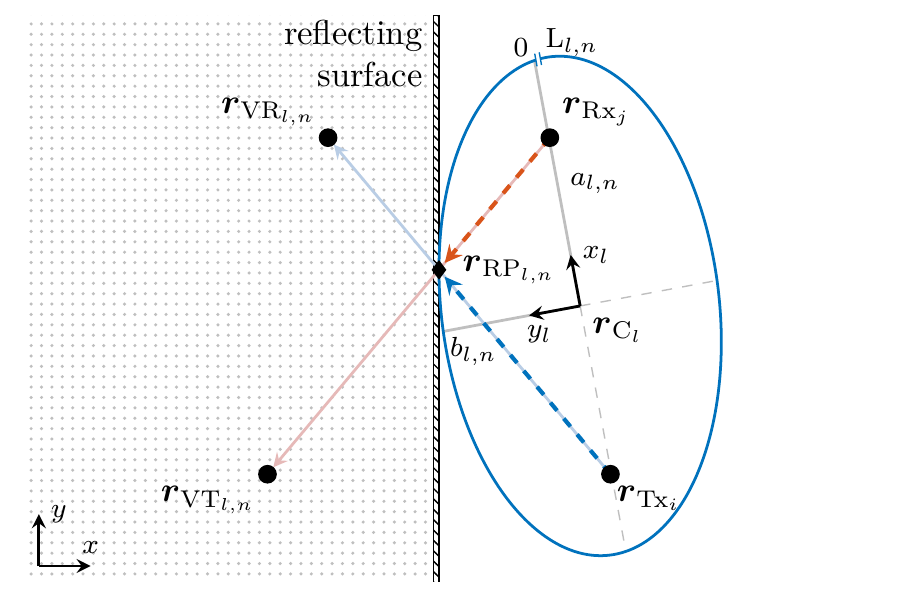}
  	\caption{Propagation path for an \acs{MPC} due to \acl{SBR} at a reflecting surface and corresponding 2D representation of the delay ellipsoid.}
  \label{fig:vt_vr_sbr} 
\end{figure}
An important special case represents \acp{MPC} caused by first-order reflections or scattering, i.e., \acp{SBR}.
Compared to \acp{MPC} caused by higher order reflections, the received power of an \ac{MPC} caused by \ac{SBR} is expected to be less attenuated~\cite{naseri2017}. 
The comparatively high received power improves the quality of the \ac{MPC} parameter estimates such as propagation delay and amplitude, cf. \eqref{eq:signal_model}.
Besides the advantages due to the received power, \acp{MPC} caused by \ac{SBR} comprise also specific geometric properties of the propagation paths, that we will outline in the following.

For \acp{SBR}, the set of virtual nodes reduces to one \ac{VT} and one \ac{VR}, so that the propagation path of the corresponding \ac{MPC} can be described by a single \ac{RP}.
As shown in Fig.~\ref{fig:vt_vr_sbr}, the respective vector between the physical node, i.e., transmitter or receiver, and the \ac{RP} is aligned with the vector between the physical and the corresponding virtual node.
That means, for the transmitting and receiving node we can equally define the unit vectors
\begin{equation}
\label{eq:VTratio}
e_j = 
\frac{\boldsymbol{r}_{\mathrm{VT}_{l,n}}-\boldsymbol{r}_{\mathrm{Rx}_{j}}}{\lVert \boldsymbol{r}_{\mathrm{VT}_{l,n}}-\boldsymbol{r}_{\mathrm{Rx}_{j}}\rVert}
 = \frac{\boldsymbol{r}_{\mathrm{RP}_{l,n}}-\boldsymbol{r}_{\mathrm{Rx}_{j}}}{\lVert \boldsymbol{r}_{\mathrm{RP}_{l,n}}-\boldsymbol{r}_{\mathrm{Rx}_{j}}\rVert},
\end{equation}
and
\begin{equation}
\label{eq:VRratio}
e_i = 
\frac{\boldsymbol{r}_{\mathrm{VR}_{l,n}}-\boldsymbol{r}_{\mathrm{Tx}_{i}}}{\lVert \boldsymbol{r}_{\mathrm{VR}_{l,n}}-\boldsymbol{r}_{\mathrm{Tx}_{i}}\rVert}
 = \frac{\boldsymbol{r}_{\mathrm{RP}_{l,n}}-\boldsymbol{r}_{\mathrm{Tx}_{i}}}{\lVert \boldsymbol{r}_{\mathrm{RP}_{l,n}}-\boldsymbol{r}_{\mathrm{Tx}_{i}}\rVert},
\end{equation}
where $\boldsymbol{r}_{\mathrm{VT}_{l,n}}$, $\boldsymbol{r}_{\mathrm{VR}_{l,n}}$, and $\boldsymbol{r}_{\mathrm{RP}_{l,n}}$ denote the locations of \ac{VT}, \ac{VR}, and \ac{RP}, respectively, for the $n$-th \ac{MPC} of network link~$l$. 
As mentioned in Sec.~\ref{sec:preVN}, the distances between the physical transmitting or receiving node and the corresponding virtual node equal the length of the propagation path as 
\begin{equation}
\label{eq:distVTVR}
\lVert \boldsymbol{r}_{\mathrm{VT}_{l,n}}-\boldsymbol{r}_{\mathrm{Rx}_{j}}\rVert
=
\lVert \boldsymbol{r}_{\mathrm{VR}_{l,n}}-\boldsymbol{r}_{\mathrm{Tx}_{j}}\rVert
=
d_{l,n}.
\end{equation}
By inserting \eqref{eq:distVTVR} into \eqref{eq:VTratio} and \eqref{eq:VRratio}, we can resolve for the locations of \ac{VT} and \ac{VR} as
\begin{equation}
\label{eq:VT}
\boldsymbol{r}_{\mathrm{VT}_{l,n}} = \frac{\boldsymbol{r}_{\mathrm{RP}_{l,n}}-\boldsymbol{r}_{\mathrm{Rx}_{j}}}{\lVert \boldsymbol{r}_{\mathrm{RP}_{l,n}}-\boldsymbol{r}_{\mathrm{Rx}_{j}}\rVert} d_{l,n} + \boldsymbol{r}_{\mathrm{Rx}_{j}},
\end{equation}
and
\begin{equation}
\label{eq:VR}
\boldsymbol{r}_{\mathrm{VR}_{l,n}} = \frac{\boldsymbol{r}_{\mathrm{RP}_{l,n}}-\boldsymbol{r}_{\mathrm{Tx}_{i}}}{\lVert \boldsymbol{r}_{\mathrm{RP}_{l,n}}-\boldsymbol{r}_{\mathrm{Tx}_{i}}\rVert} d_{l,n} + \boldsymbol{r}_{\mathrm{Tx}_{i}}.
\end{equation}
%Therewith, we have a unique expression of the locations of \ac{VT} and \ac{VR} given the locations of the \ac{RP} and the physical nodes

Besides these unique expressions for the locations of \ac{VT} and \ac{VR}, the propagation delay of an \ac{MPC} caused by \ac{SBR} allows to constrain the possible locations of the corresponding \ac{RP}.
Specifically, this means that the possible locations of an \ac{RP} coincide with the locations on the surface of an ellipsoid, each of which equally causes the propagation delay~$\tau_{l,n}$~\cite{walter2014,walter2020}.
For reflections from surfaces orthogonal to the axis between the transmitting and receiving nodes, e.g., vertical walls of buildings, the possible locations are reduced to the circumference of an ellipse, as shown in Fig.~\ref{fig:vt_vr_sbr}.
Since we expect the majority of \acp{MPC} to originate from reflections off surrounding walls, we adopt this \ac{2D} representation of an ellipse.
We therefore also assume that all transmitting and receiving nodes are located at the same height, i.e., in the same plane.
Note that all vectors and vector operations in the remainder of this paper are considered in \ac{2D}.
In general, the shape of an ellipse is determined by its principal axes. 
These include the semi-major and the semi-minor axis, which are defined for the $n$-th~\ac{MPC} of network link~$l$ as
\begin{equation}
\label{eq:ell_a}
{a}_{l,n} = \frac{d_{l,n}}{2},
\end{equation}
and
\begin{equation}
\label{eq:ell_b}
{b}_{l,n} = \frac{\sqrt{d_{l,n}^2 - d_{l}^2}}{2},
\end{equation}
with path lengths $d_{l,n}$ and $d_{l}$ as given in \eqref{eq:prop_dist} and \eqref{eq:prop_dist_LOS}, see Fig.~\ref{fig:vt_vr_sbr}.
The center of the ellipse is given by
\begin{equation}
\label{eq:ell_center}
\boldsymbol{r}_{\mathrm{C}_{l}} = \frac{\boldsymbol{r}_{\mathrm{Rx}_{j}}+\boldsymbol{r}_{\mathrm{Tx}_{i}}}{2}.
%\frac{\boldsymbol{r}_{\mathrm{Rx}_{j}}-\boldsymbol{r}_{\mathrm{Tx}_{i}}}{2} + \boldsymbol{r}_{\mathrm{Tx}_{i}}.
\end{equation}
In a local coordinate system with the center of the ellipse $\boldsymbol{r}_{\mathrm{C}_{l}}$ at the origin and the semi-major and semi-minor axes along $x_{l}$-axis and $y_{l}$-axis, see Fig.~\ref{fig:vt_vr_sbr}, the ellipse can be expressed analytically by
\begin{equation}
\label{eq:ell_eq}
\frac{{x_{l}}^2}{{a}_{l,n}^2} + \frac{{y_{l}}^2}{{b}_{l,n}^2} = 1,
\end{equation}
and, equivalently, by the parametric representation
\begin{equation}
\label{eq:ell_eq_polar}
\begin{bmatrix}
x_{l} \\
y_{l}
\end{bmatrix}
= 
\begin{bmatrix}
{a}_{l,n} \, \cos{\theta} \\
{b}_{l,n} \, \sin{\theta}
\end{bmatrix},
\end{equation}
with the ellipse parameter $\theta \in \left[ 0,2 \pi\right]$.
Note that the local ellipse-centric coordinate system is defined individually for each network link~$l$, but applies to all \acp{MPC} and corresponding delay ellipses of that link.
Unlike a circle, there is no simple relation between the ellipse parameter with different semi-minor and semi-major axes and the arc length~\cite{norklit1998}.
In order to describe the distribution of possible locations of an \ac{RP} along the elliptic arc, we need to define the arc length as a function of the ellipse parameter~\cite{norklit1998}, as
\begin{equation}
\label{eq:ell_arc}
s_{l,n}(\theta) = 
{a}_{l,n} \int_0^{\theta}\sqrt{1-\epsilon_{l,n}^2 \, \cos^2\vartheta}\,\text{d}\vartheta,
\end{equation}
with the numerical eccentricity
\begin{equation}
\label{eq:ell_arc_ecc}
\epsilon_{l,n} = \sqrt{1-\frac{{b}_{l,n}^2}{{a}_{l,n}^2}}.
%\epsilon_{l,n} = \frac{\sqrt{{a}_{l,n}^2-{b}_{l,n}^2}}{{a}_{l,n}}.
\end{equation}
Therewith, we can further calculate the circumference of the ellipse as
\begin{equation}
\label{eq:ell_circum}
\text{L}_{l,n} = 4\,{a}_{l,n} \int_0^{\pi/2}\sqrt{1-\epsilon_{l,n}^2 \, \cos^2\vartheta}\,\text{d}\vartheta.
\end{equation}
With $E(\epsilon_{l,n}) = \int_0^{\pi/2} \sqrt{1-\epsilon_{l,n}^2\cos^2\vartheta}\,\text{d}\vartheta$ as the complete elliptic integral of the second kind, \eqref{eq:ell_circum} reduces to
\begin{equation}
\label{eq:ell_circum_short}
\text{L}_{l,n} = 4\,{a}_{l,n} E(\epsilon_{l,n}).
\end{equation}

Finally, we want to express the local ellipse-centric coordinates $\boldsymbol{x}_{l}$ in the coordinate system used for the \ac{MDFL} network.
Thus, the local coordinates are first rotated by the rotation matrix $\boldsymbol{R}_{l}$ and translated by the center of the ellipse~$\boldsymbol{r}_{\mathrm{C}_{l}}$ as 
\begin{equation}
\label{eq:cart_ell}
%\boldsymbol{x} = \boldsymbol{R}_{l} \, \boldsymbol{x}_{l,n} - \boldsymbol{r}_{\mathrm{C}_{l}}.
\boldsymbol{x} = \boldsymbol{R}_{l} \, \boldsymbol{x}_{l} + \boldsymbol{r}_{\mathrm{C}_{l}}.
\end{equation}
Here, we consider the rotation of the local coordinates in the $x\,y$-plane with respect to the $x$-axis about the origin of the coordinate system used for \ac{MDFL}.
Therefore, the rotation matrix is expressed as
\begin{equation}
\label{eq:cart_ell_Rot}
\boldsymbol{R}_{l} = 
\begin{bmatrix}
\cos{\alpha_{l}} & -\sin{\alpha_{l}}  \\
\sin{\alpha_{l}} &  \cos{\alpha_{l}}  \\
\end{bmatrix},
%\begin{bmatrix}
%\cos{\alpha_{l}} & -\sin{\alpha_{l}} & 0  \\
%\sin{\alpha_{l}} &  \cos{\alpha_{l}} & 0  \\
%0 & 0 & 1   
%\end{bmatrix},
\end{equation}
with rotation angle $\alpha_{l}$. 
The angle determines the rotation between the $x$-axis and the $x_{l}$-axis of the local coordinate system. With the unit vectors $\boldsymbol{e}_x$ in $x$-direction and 
${\boldsymbol{e}_{x_{l}} =\left(\boldsymbol{r}_{\mathrm{Rx}_{j}}-\boldsymbol{r}_{\mathrm{Tx}_{i}}\right)/\lVert \boldsymbol{r}_{\mathrm{Rx}_{j}}-\boldsymbol{r}_{\mathrm{Tx}_{i}} \rVert}$ in $x_{l}$-direction the rotation angle can be calculated by
\begin{equation}
\label{eq:cart_ell_alpha}
\alpha_{l} = \text{sgn}_l\, \arccos{\left( \boldsymbol{e}_x^T \, \boldsymbol{e}_{x_{l}}\right)},
%\alpha_{l} = \sign \left( \det\left[ \boldsymbol{e}_{x_{l,n}} ,\boldsymbol{e}_x \right]\right) \arccos{\left( \boldsymbol{e}_{x_{l,n}}^T \, \boldsymbol{e}_x\right)},
%\alpha_{l} = \arccos{\left( \boldsymbol{e}_x^T \, \frac{\boldsymbol{r}_{\mathrm{Rx}_{j}}-\boldsymbol{r}_{\mathrm{Tx}_{i}}}{\lVert \boldsymbol{r}_{\mathrm{Rx}_{j}}-\boldsymbol{r}_{\mathrm{Tx}_{i}} \rVert}\right)},
%\alpha_{l,n} = \arccos{\left( \frac{\left(\boldsymbol{r}_{\mathrm{Rx}_{j}}-\boldsymbol{r}_{\mathrm{Tx}_{i}}\right)}{\lVert \boldsymbol{r}_{\mathrm{Rx}_{j}}-\boldsymbol{r}_{\mathrm{Tx}_{i}} \rVert}^T \, \boldsymbol{e}_x\right)},
\end{equation}
where the direction of the rotation is determined by the sign of the determinant of the unit vectors, i.e., ${\text{sgn}_l = \sign\left(\det \left[\boldsymbol{e}_x , \boldsymbol{e}_{x_{l}}\right]\right)}$.

%Particularly, the possible locations of an \ac{RP} are defined by the surface of an ellipsoid
%Moreover, the possible locations of the \ac{RP} of an \ac{MPC} due to \ac{SBR} are geometrically constrained by the length of the propagation path, i.e., the propagation delay as defined in~\eqref{eq:prop_dist}.
%Therefore, the \ac{RP} has to be located on the surface of an ellipsoid with foci equal to the locations of transmitting and receiving node~\cite{walter2020}.
%For clarity of presentation and avoiding complexity, we reduce the three-dimensional representation to a two-dimensional representation, i.e., the possible locations of the \ac{RP} are described by an ellipse in the remainder of this paper.
%\section{Empirical Modeling of Body Occupancy}
%\section{Empirical Modeling of User Induced Fading of Multipath Components}
% \section{Empirical Occupancy Model for MPCs}
\begin{figure*}[ht!]
	\centering
   	\subfloat[\label{fig:meas_hangar}Setup~I: controlled apron environment]{%
	   	\includegraphics[width=2.78in]{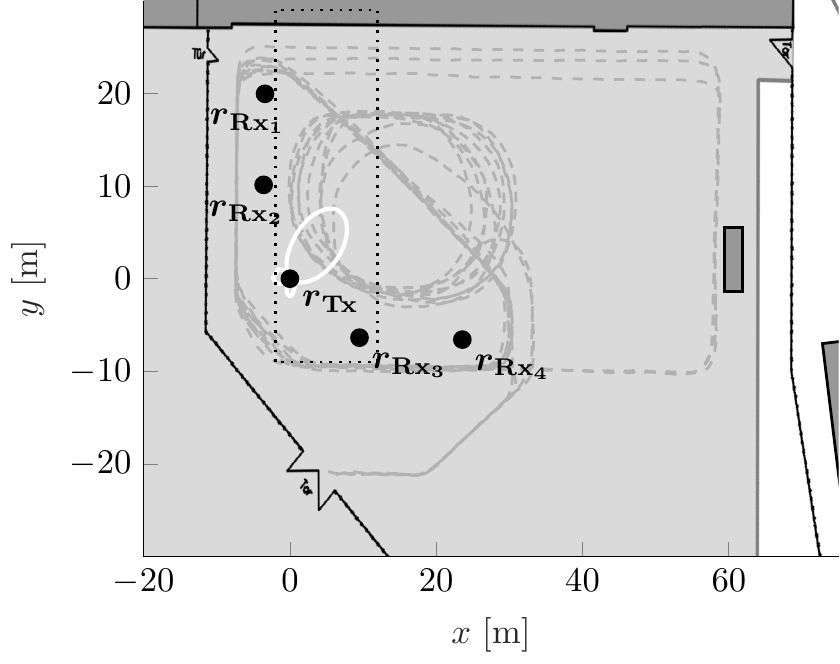}}
	\centering
	\hspace*{-0.06in}
   	\subfloat[\label{fig:meas_techlab}Setup~II: typical urban intersection environment]{%
	   	\includegraphics[width=2.78in]{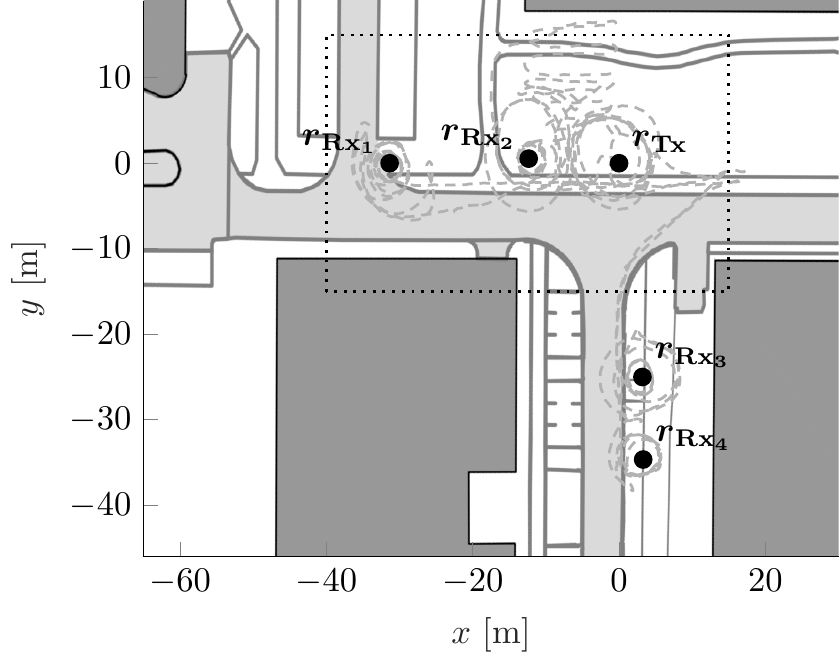}}
	\centering
	\hspace*{-0.00in}
   	\subfloat[\label{fig:meas_holodeck}Setup~III: indoor environment]{%
	   	\includegraphics[width=1.6in]{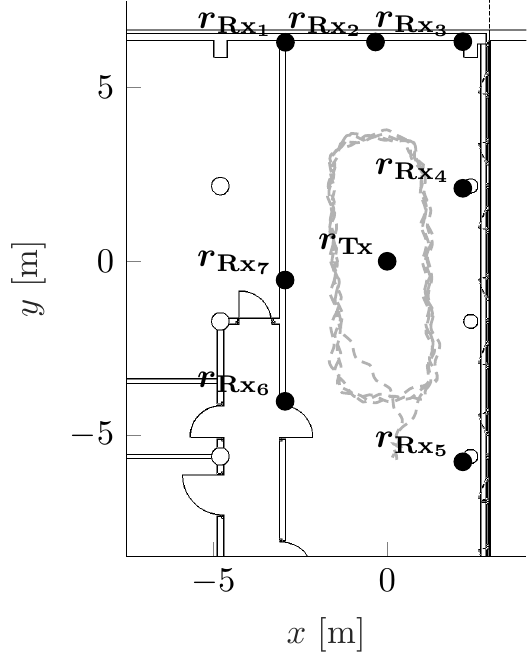}}
	\setlength{\belowcaptionskip}{-5pt}
  	\caption{Measurement setups showing transmitter and receiver locations~(\mytikzNODE), user trajectories~(\mytikzLineDashGray), and distinct reflection surfaces, e.g. building walls~(\mytikzLineSolidThick).
  	 In (a), the shape of the directional antenna pattern around $\boldsymbol{r}_{\text{Tx}}$ is highlighted in light gray. The dotted rectangles in (a) and (b) refer to the map sections used for Fig.~\ref{fig:mpc_example} and Fig.~\ref{fig:rp_pmf_meas_env}, respectively.
  }
  \vspace{0.1cm}
  \label{fig:meas} 
\end{figure*}

\section{Empirical Fading Model for MPCs}
\label{sec:model}
%\begin{savenotes}
\begin{table*}[t]
    \centering
    \caption{Overview measurement setups and parameters}
%    \comment{3.9936 GHz}{high precision compared to 5.2 GHz. Why necessary? same for wavelengths below.}
    \label{tab:meas_overview}
    \begin{tabular}{lcccc}
        \toprule
         & & \multicolumn{1}{p{1.7in}}{\centering \textbf{Setup I}} & \multicolumn{1}{p{1.7in}}{\centering \textbf{Setup II}} & \multicolumn{1}{p{1.7in}}{\centering \textbf{Setup III}} \\
  %       & & \multicolumn{1}{p{0.6in}}{\centering \textbf{Setup I}} & \multicolumn{1}{p{0.6in}}{\centering \textbf{Setup I}} & \multicolumn{1}{p{0.6in}}{\centering \textbf{Setup III}} \\
        \midrule \addlinespace
        Measurement system      &     &   \multicolumn{2}{c}{Medav DLR-RUSK wideband channel sounder}  & DecaWave UWB DW 1000~\cite{dw1000}\\   \addlinespace
        Transmit signal                &         &  \multicolumn{2}{c}{right hand circular polarized multitone}   & IEEE 802.15.4-2011 \\ \addlinespace
        Center frequency        &  $f_{\text{c}}$ &  \multicolumn{2}{c}{\SI{5.2}{\GHz}} & \SI{3.9936}{\GHz} \\ \addlinespace
        Wavelength        &  $\lambda$ &   \multicolumn{2}{c}{\SI{5.77}{\cm}}  & \SI{7.51}{\cm}
%        Wavelength        &  $\lambda$ &   \multicolumn{2}{c}{\SI{5.765}{\cm}}  & \SI{7.507}{\cm}
       % \footnote{Due to the high bandwidth, the wavelength at the band edges differ.}
        \\ \addlinespace
        Bandwidth               & $B$   &   \multicolumn{2}{c}{\SI{120}{\MHz}}  & \SI{499.2}{\MHz} \\ \addlinespace
        Update time             & $T_\text{g}$ & \multicolumn{2}{c}{\SI{2.048}{\ms}}  & $\approx$ \SI{0.3}{\s} \\ \addlinespace
        Sampling period         & $T_\text{p}$ & \multicolumn{2}{c}{\SI{3.2}{\micro\s}}  & \SI{0.11}{\micro\s} \\ \addlinespace
        Transmit power          & $P_\text{Tx}$ & \multicolumn{2}{c}{\SI{37}{\dBm}}  &  $\leq$\SI{12.67}{\dBm} \\ \addlinespace
        Measurement setup       &       &   \multicolumn{2}{c}{$N_{\text{Tx}} = 1$, $N_{\text{Rx}} = 4$}  & $N_{\text{Tx}} = 1$, $N_{\text{Rx}} = 7$  \\ 
                                &       &   \multicolumn{2}{c}{SIMO $1\times4$}  & individually addressed, round-robin  \\ \addlinespace                
Transmit antenna        & & \SI{9}{\dBi} directional~\cite{hubersuhner1399_17_0210} & \SI{8}{\dBi} toroidal, omni-directional~\cite{hubersuhner1399_17_0111}  & \SI{0}{\dBi} omni-dirctional~\cite{dw1000,partronACS5200HFAUWB}  \\ \addlinespace 
        Receive antennas        & & \SI{8}{\dBi} toroidal, omni-directional~\cite{hubersuhner1399_17_0111}& \SI{8}{\dBi} toroidal, omni-directional~\cite{hubersuhner1399_17_0111}  & \SI{0}{\dBi} omni-dirctional~\cite{dw1000,partronACS5200HFAUWB}  \\ \addlinespace 
        Reference system  &  &tachymeter & GNSS & optical motion capture system \\ \addlinespace
        Environment           & &  outdoors & outdoors & indoors \\ \addlinespace
        User types  & &  pedestrian &  pedestrian & pedestrian  \\
                           &          &   bike & & \\
                            &         &   car &  &  \\ \addlinespace
                                     
    \bottomrule
    \end{tabular}
\end{table*}
%\end{savenotes}
%\highlight{The objective of}  \comment{is the estimation of the user location based on user-induced changes in the received signal power.}{sounds heavy, indirect. please rephrase more direct, e.g. MDFL aims at estimating the user location ... }
\ac{MDFL} systems aim to estimate the user location based on user-induced changes in the received signal power.
Therefore, the measurement model that relates the measured changes in the received signal power to the user location is essential for any \ac{MDFL} system.
The performance of the localization system depends on the robustness, complexity, accuracy, and validity of the underlying model.
In this section, we therefore derive and validate an empirical model describing the user-induced variations in the received power of \acp{MPC}, based on measurement data. 
\subsection{Data Collection}
\label{sec:model_measurements}
For investigating the impact of users on the power of \acp{MPC}, we have collected an extensive set of wideband and \ac{UWB} measurement data in different multipath environments, both indoors and outdoors.
Accounting for a variety of different influencing parameters, we consider data from three different measurement campaigns.
Thereby, the sets contain data from different configurations, different environments, and different types of users. Table~\ref{tab:meas_overview} provides a detailed overview of the settings and parameters used during the individual measurement campaigns.
The corresponding network configurations and geometric setups are illustrated in Fig.~\ref{fig:meas}.
In the following, we briefly describe the individual measurement campaigns.

\subsubsection{Setup I}
The measurement campaign was conducted outdoors on a flight apron providing a largely controlled environment.
Using the MEDAV RUSK-DLR wideband channel sounder in $1\times4$ \ac{SIMO} mode, we measured repeatedly \ac{CIR} snapshots of four radio channels.
The measurement setup consisted of one transmitting and four receiving antennas statically arranged as shown in Fig.~\ref{fig:meas_hangar}.
All antennas were placed \SI{1}{\m} above the ground.
We used four similar toroidal, omni-directional receive antennas.
To avoid an elevation of the noise floor due to limits of the dynamic range, we used a directional transmit antenna in order to reduce the received signal power of the \ac{LoS} component in each channel~\cite{schmidhammer2019sensors}.
The orientation of the transmit antenna is indicated by the light gray radiation pattern in Fig.~\ref{fig:meas_hangar}.
For inducing variations on the received signal power, different users were moving individually within the measurement environment.
In this campaign, pedestrians, bicyclists, and cars were considered as users.
%We recorded the ground truth of the moving users using a tachymeter together with a high precision reflector prism.
By using a tachymeter in conjunction with a high-precision reflector prism, we recorded the ground truth of the moving users.
%Using an elevated tachymeter together with a high precision reflector prism, the ground truth of the moving users was recorded.
The reflector prism was attached to a helmet worn by the pedestrians and bicyclists, and was mounted centrally on the roof of the cars, respectively. 
Fig.~\ref{fig:meas_hangar} illustrates the trajectories of the pedestrian user as an example.
The trajectories of the bicyclists and cars were comparable.
In total, we collected more than \SI{2100000}{} \ac{CIR} snapshots for pedestrian users, and more than \SI{1300000}{} each for bicycle and car users.

\subsubsection{Setup II}
The measurement campaign was also conducted outdoors, but at an intersection on the DLR campus.
The setup was openly accessible, i.e., the environment was less controlled.
In addition, the propagation environment, with nearby buildings, parked cars, and vegetation, resembles that of a typical urban intersection.
Similarly to Setup~I, the MEDAV RUSK-DLR wideband channel sounder was used in $1\times4$ \ac{SIMO} mode collecting \ac{CIR} snapshots of four channels. 
The measurement setup is shown to scale in Fig.~\ref{fig:meas_techlab}.
Again, the antennas were placed \SI{1}{\m} above the ground.
Here, however, we used the same toroidal, omni-directional antennas at the transmitter and receivers to account for \acp{MPC} from all directions.
In this setup, one moving pedestrian is considered, which induces variations to the received signal.
Ground truth was recorded using a u-blox F9R \ac{GNSS} receiver connected to an antenna attached to the pedestrian's helmet, similar to the prism in Setup~I.
Using real-time differential \ac{GNSS} corrections, the receiver computed a \ac{GNSS} real-time kinematic solution.
Due to the fair open-sky environment, a sub-decimeter accuracy can be expected.
The trajectory of the pedestrian is also shown in Fig.~\ref{fig:meas_techlab}.
In total, more than \SI{182000}{} \ac{CIR} snapshots were recorded.%
\footnote{
%\highlight{The measurement campaign was conducted for the project VIDETEC funded by the Federal Ministry of Transport and Digital Infrastructure (BMVI) as part of the mFUND Mobility Pact 4.0.}
Note that parts of the measurement data will be made publicly available through the open data platform mCLOUD~\cite{mCloud}.}

\subsubsection{Setup III} 
The measurement campaign was conducted indoors in a controlled environment.
Using an \ac{UWB} measurement system, based on the commercial off-the-shelf DecaWave DW 1000 chip, we collected the \ac{CIR} snapshots for seven channels.
Thereby, the measurement setup consisted of one transmitting and seven receiving \ac{UWB} nodes as shown in Fig.~\ref{fig:meas_holodeck}.
For measuring the \ac{CIR}, the receiving nodes were individually addressed by the transmitting node in a round-robin manner.
All nodes were mounted \SI{1}{\m} above the ground.
In this campaign, one moving pedestrian was considered, where the ground truth was recorded using a Vicon high-precision optical motion capture system.
Therefore, a reflector was attached on top of the head of the pedestrian, which was tracked by the motion capture system.
The Vicon motion capture system is capable of tracking the motion of the reflector with a sub-centimeter accuracy.
The trajectory is shown in Fig.~\ref{fig:meas_holodeck}.
In total, we collected more than \SI{2800}{} \ac{CIR} snapshots.

\subsection{Parameter Estimation}
\label{sec:model_parameter_est}
For each channel measured during the measurement campaigns presented above, we must first determine the specific propagation effects of the static environment. 
In each campaign, we therefore collected additional \ac{CIR} samples over a period during which the environment was devoid of any user.
%of about $T_\text{ini} \approx \SI{1}{\min}$
For each of these measured \acp{CIR}, we can now determine the signal parameter for $N_l$ separable \acp{MPC}, i.e., the amplitude and delay values as defined in \eqref{eq:signal_model}, using maximum likelihood estimation.
Specifically, we use the \ac{SAGE} algorithm for parameter estimation ~\cite{fleury1999}.
Note that the measured channels of each campaign correspond to network links of an equivalent \ac{MDFL} system, which allows the use of the same notation as introduced in Sec.~\ref{sec:preNet}.

Subsequently, we average the estimated signal parameters over time and obtain a mean amplitude
$\bar{\alpha}_{l,n}$
and a mean delay value
$\bar{\tau}_{l,n}$
for each \ac{MPC} $n$ of channel $l$.
The mean amplitude values allow to calculate the power of the \acp{MPC} for the idle channel.
In logarithmic domain, the power of the $n$-th \ac{MPC} of channel $l$ is
\begin{equation}
\label{eq:amplitude_MeanPower}
\bar{\gamma}_{l,n} = 20 \log_{10}\left|\bar{\alpha}_{l,n} \right|,
\end{equation}
which can be used as reference power level for measuring user-induced power changes.

For this purpose, we determine the amplitude values for all \acp{MPC} of each channel.
Given the mean delay $\bar{\tau}_{l,n}$ for the $n$-th \ac{MPC} of channel $l$, we estimate the amplitude using
\begin{equation}
\label{eq:amplitude_est}
\hat{\alpha}_{l,n} = {\alpha}(\bar{\tau}_{l,n}) = \int_0^{T_\text{p}} \left(y_{l,n}^{\mathrm{res}}(t)\right)^{*} s_l(t-\bar{\tau}_{l,n}) \, d t,
\end{equation}
that is, the projection of the residuum signal 
$y_{l,n}^{\mathrm{res}}(t)$
onto the unit transmit signal $s_l(t)$~\cite{meissner2014,schmidhammer2021_awpl}.
Adjusting the received signal for all \acp{MPC} up to the $(n-1)$-th, the residuum signal is calculated as
\begin{equation}
\label{eq:amplitude_est_res}
y_{l,n}^{\mathrm{res}}(t) =y_{l}(t)- \sum_{n'=1}^{n-1}\hat{\alpha}_{l,n'}s_l(t-\bar{\tau}_{l,n'}).
\end{equation}
Given the amplitude estimate, we can then express the measured power in logarithmic domain as
\begin{equation}
\label{eq:amplitude_MeasPower}
\hat{\gamma}_{l,n} = 20 \log_{10}\left|\hat{\alpha}_{l,n} \right|.
\end{equation}
By subtracting the reference power level in \eqref{eq:amplitude_MeanPower} from the measured power in \eqref{eq:amplitude_MeasPower} we can finally express the user-induced power changes as
\begin{equation}
\label{eq:amplitude_ChangePower}
{z}_{l,n} = \hat{\gamma}_{l,n}-\bar{\gamma}_{l,n} = 20 \log_{10}\frac{\left|\hat{\alpha}_{l,n} \right|}{\left|\bar{\alpha}_{l,n} \right|}.
\end{equation}

In order to describe the location dependent impact of a user on the power of an \ac{MPC}, we must next determine the actual propagation path of the \ac{MPC}.
Therefore, for each measurement setup, we have accurately measured the locations of distinct reflective objects in the propagation environment, including walls, lampposts, and fences, as well as vegetation and furniture. 
Using additional precise floor plan information, we obtain a detailed map of possible locations of \acp{RP}.
Applying \eqref{eq:ell_eq_polar} and \eqref{eq:cart_ell}, we then construct the delay ellipse for $\bar{\tau}_{l,n}$, as shown exemplary in Fig.~\ref{fig:vt_vr_sbr}.
By calculating the tangent or intersection points of the ellipse within the previously defined map, we determine potential \acp{RP} of the \ac{MPC}.
%Finally, we manually identify the true \ac{RP} based on ground truth information of the mobile user together with the measured power changes.
%\TODO{Maybe here a reference to the Bayesian RP mapping algorithm in the next section.}
In this way, we determine all uniquely resolvable \acp{RP} for the \acp{MPC} of each measured channel.
As examples, Fig.~\ref{fig:meas_hangar_RP} and Fig.~\ref{fig:meas_holodeck_RP} illustrate the resulting propagation paths for Setup~I and Setup~III, respectively.
Please note that while mandatory for the empirical modeling of this section, the manual determination of \acp{RP} is extremely cumbersome and can only be performed in post-processing.
%Please note that the manual determination of \acp{RP}, which is mandatory for the empirical modeling of this section, is extremely cumbersome and can only be applied in post-processing.
Flexible adaptation to a changing environment is thus basically impossible, which further motivates for an automated \ac{RP} estimation and mapping approach, as presented in Sec.~\ref{sec:bayes_mapping}.

\subsection{Empirical Exponential Model}
\label{sec:model_exp_model}
\label{sec:empirical_model}
\begin{figure*}[t!]
	\centering
   	\subfloat[\label{fig:model_raw_data} Power change and fitted exponential fading model]{%
	   	\includegraphics[width=2.5in]{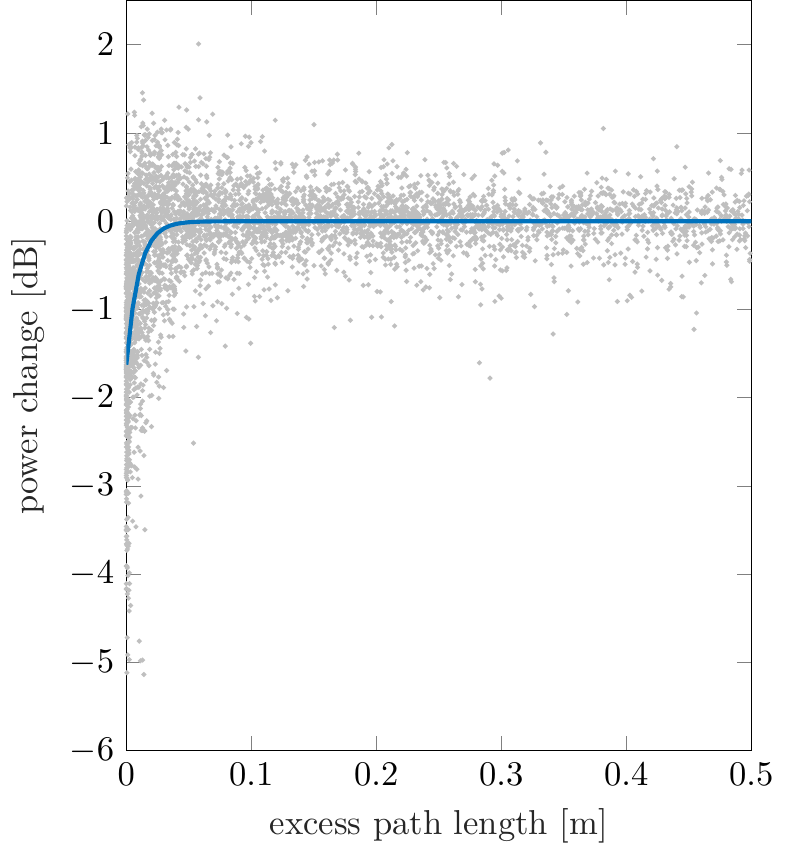}}
	\centering
	\hfill
   	\subfloat[\label{fig:model_residual} Corresponding residual power and standard deviation]{%
	   	\includegraphics[width=2.5in]{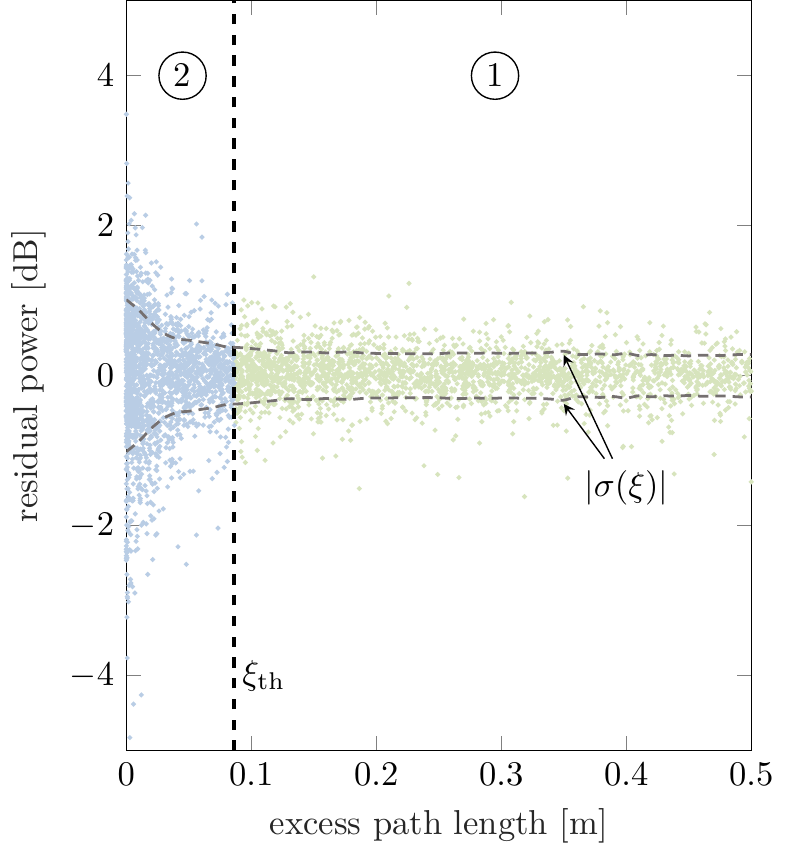}}
	\centering
	\hfill
   	\subfloat[\label{fig:model_pdf} \acp{PDF} of residual power]{%
	   	\includegraphics[height=2.8in]{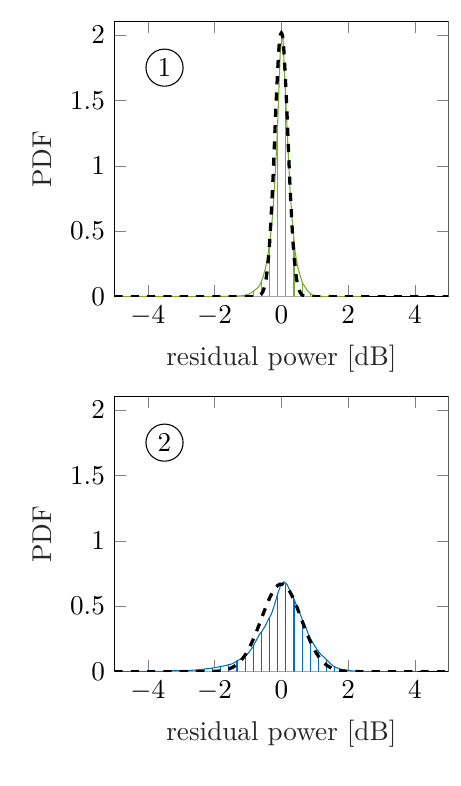}}
	\centering
  	\caption{Measurement data for pedestrian user in Setup~I: 
  	(a) measured power change values and fitted exponential fading model;
    (b)~corresponding residual power values along with standard deviation (enveloping gray dashed lines), and Fresnel-motivated threshold for excess path length (vertical dashed line);
    (c)~\acp{PDF} of residuals for excess path length above and below the threshold ($\mytikzONE$ and $\mytikzTWO$), where dashed lines in black indicate the fitted normal distribution with $\sigma_1 = \SI{0.2813}{\dB}$ and $\sigma_2 = \SI{0.7957}{\dB}$ (cf. Table~\ref{tab:model_noise}).}
%  	large-scale fading model of pedestrian for Setup I are given in (a). Residual values are given in (b), with corresponding standard deviation as gray dashed line and Fresnel threshold as vertical black dashed line. The PDFs for excess path lengths smaller and larger than the threshold are shown in (c), where the dashed lines indicate the fitted normal distribution with $\sigma_1 = 0.7957$ and $\sigma_2 = 0.2813$ (cf. Table~\ref{tab:model_noise}).}
  \label{fig:model_meas} 
\end{figure*}
After collecting and processing the measurement data, we can now investigate the location dependent impact of a user on the power of an \ac{MPC}.
Therefore, we must first describe the relation between the user location $\boldsymbol{r}$ and the propagation path of an \ac{MPC}.
A common measure that relates the user's location to a signal propagation path is the excess path length~\cite{guo2015,kaltiokallio2021,schmidhammer2021_awpl}.
For calculating the excess path length of an \ac{MPC}, we need to geometrically decompose the propagation path by combining successive pairs of virtual and physical nodes (see Sec.~\ref{sec:preVN}).
Thus, the excess path length of an \ac{MPC} caused by \ac{SBR} is calculated as
\begin{equation}
\label{eq:excess_TX}
\xi_{l,n}^{\text{Tx}}(\boldsymbol{r}) =
\Vert \boldsymbol{r}_{\mathrm{Tx}_{i}} - \boldsymbol{r} \Vert + \Vert \boldsymbol{r}_{\mathrm{VR}_{l,n}} - \boldsymbol{r} \Vert - d_{l,n},
\end{equation}
for the link between physical transmitter and \ac{VR}, see blue line in Fig.~\ref{fig:vt_vr_sbr}, and as
\begin{equation}
\label{eq:excess_RX}
\xi_{l,n}^{\text{Rx}}(\boldsymbol{r}) =
\Vert \boldsymbol{r}_{\mathrm{VT}_{l,n}} - \boldsymbol{r} \Vert + \Vert \boldsymbol{r}_{\mathrm{Rx}_{j}} - \boldsymbol{r} \Vert - d_{l,n},
\end{equation}
for the link between \ac{VT} and physical receiver, see red line in~Fig.~\ref{fig:vt_vr_sbr}.
Following~\cite{schmidhammer2021_awpl}, we approximate the user impact for the two pairs of nodes individually by the empirical exponential model~(cf.~\cite{guo2015}).
Therewith, the user location dependent power changes of an \ac{MPC} are modeled as 
\begin{equation}
\label{eq:exp_model_sumMPC}
f_{l,n}(\boldsymbol{r}) =  \phi_{l,n} \left(e^{-\xi_{l,n}^{\text{Tx}}(\boldsymbol{r})/\kappa_{l,n}} + e^{-\xi_{l,n}^{\text{Rx}}(\boldsymbol{r})/\kappa_{l,n}} \right),
%f_{l,n}(\boldsymbol{r}) =  \phi_{l,n} e^{-\delta_{l,n}^{\text{Tx}}(\boldsymbol{r})/\kappa_{l,n}} + \phi_{l,n} e^{-\delta_{l,n}^{\text{Rx}}(\boldsymbol{r})/\kappa_{l,n}} ,
\end{equation}
with $\phi_{l,n}$ and $\kappa_{l,n}$ being the maximum modeled power change in \SI{}{\dB} and the spatial decay rate. 
Note that in practice the model parameter can be determined individually for each link and each \ac{MPC}.
For the following empirical derivation and evaluation of the model parameter, however, we use the measured power changes of all \acp{MPC} from all channels to obtain sufficient statistics.
%For the subsequent evaluation, however, we use the measured power changes of all \acp{MPC} from all channels.
Accordingly, Fig.~\ref{fig:model_raw_data} illustrates the power changes measured for all \acp{MPC} of each channel, induced by the pedestrian in Setup~I.
Additionally, the figure highlights the exponential fading model of \eqref{eq:exp_model_sumMPC} with fitted model parameters.
Note that, for illustration purposes, we use the minimum excess path length for the abscissa of Fig.~\ref{fig:model_raw_data} and Fig.~\ref{fig:model_residual}.
For \ac{SBR}, the minimum excess path length is defined as
\begin{equation}
\label{eq:excess_min}
\xi_{l,n}^{\text{min}}(\boldsymbol{r}) = \min (\,\xi_{l,n}^{\text{Tx}}(\boldsymbol{r}),\,\xi_{l,n}^{\text{Rx}}(\boldsymbol{r})\,),
\end{equation}
with excess path lengths $\xi_{l,n}^{\text{Tx}}(\boldsymbol{r})$ and $\xi_{l,n}^{\text{Rx}}(\boldsymbol{r})$ defined in \eqref{eq:excess_TX} and \eqref{eq:excess_RX}.
% The propagation paths of all \acp{MPC} considered in Setup~I and used for Fig.~\ref{fig:model_meas} are illustrated in Fig.~\ref{fig:meas_hangar_RP}.
See Fig.~\ref{fig:meas_hangar_RP} for details on the propagation paths of Setup~I used for Fig.~\ref{fig:model_meas}.
Finally, Table~\ref{tab:model_exp} provides a summary of the fitted model parameter for all setups. 

Originating from the lognormal shadowing model~\cite{coulson1998}, the measured power in logarithmic domain is very often modeled by a Gaussian random variable, e.g., in~\cite{nannuru2013,guo2015,kaltiokallio2021}.
Correspondingly, the measured power changes defined in~\eqref{eq:amplitude_ChangePower} are simply modeled as
\begin{equation}
\label{eq:residual_model_unary}
    {z}_{l,n} = f_{l,n}(\boldsymbol{r}) + w_{l,n}^{}, 
\end{equation}
with the zero-mean Gaussian noise $w_{l,n}^{} \sim \mathcal{N}(0,\bar{\sigma}_{l,n}^2)$.
For Setup~I, the residual values, i.e., the difference between the fitted exponential fading model \eqref{eq:exp_model_sumMPC} and the measured power changes \eqref{eq:amplitude_ChangePower}, are illustrated in Fig.~\ref{fig:model_residual}.
Note that unlike the work in~\cite{guo2015}, we could not statistically confirm that the residuals follow a Gaussian distribution with zero mean, although the overall shape of the residuals would be comparable (cf. Fig.~7 in \cite{guo2015}).
Instead, we can clearly observe a location dependent variance as shown in Fig.~\ref{fig:model_residual}. 
In particular, the variance of the residuals increases with decreasing excess path length, i.e., the closer the user is to the propagation path, the higher the variance of the measured \acp{MPC}.
Accounting for the location dependence of the variance, we modify~\eqref{eq:residual_model_unary} as
\begin{equation}
\label{eq:residual_model_r_dep}
    {z}_{l,n} = f_{l,n}(\boldsymbol{r}) + w_{l,n}^{}(\boldsymbol{r}), 
\end{equation}
with the location dependent and zero-mean Gaussian distributed noise term $w_{l,n}^{}(\boldsymbol{r}) \sim \mathcal{N}(0,\sigma_{l,n}^2(\boldsymbol{r}))$.
\begin{table}[t]
    \centering
    \renewcommand{\arraystretch}{1.2} % Default value: 1
    \caption{Fitted model parameters}
    \label{tab:model_exp}
    \begin{tabular}{lccc}
        \toprule
        \multirow{2}{*}{ } & \multirow{2}{*}{ \textbf{User type}} & \multicolumn{2}{c}{\textbf{Model parameter}} \\
        \cmidrule(lr){3-4} 
%        \cmidrule(lr){3-5} \cmidrule(lr){6-6} \cmidrule(lr){7-7} 
        &  & $\phi$ [\SI{}{\dB}] & $\kappa$ [\SI{}{\m}]\\ 
        \midrule \addlinespace
        \textbf{Setup I} & \multicolumn{1}{p{1.8cm}}{\centering {pedestrian}} & \multicolumn{1}{p{1.5cm}}{\centering {1.6255}} &  \multicolumn{1}{p{1.5cm}}{\centering {0.0100}} \\  
         & bike & 3.2809 & 0.0193 \\  
         & car & 8.7315 & 0.1360 \\   \addlinespace
        \textbf{Setup II} & pedestrian & 4.5760 & 0.0089 \\   \addlinespace
        \multicolumn{1}{p{1.8cm}}{ {\textbf{Setup III}}} & pedestrian & 3.1267 & 0.0190 \\   \addlinespace
        \bottomrule
    \end{tabular}
\end{table}
Consistent with the work in \cite{savazzi2014}~and~\cite{rampa2015} defining detection and sensitivity areas near \ac{LoS} paths, we distinguish between areas near and far from the propagation path of an \ac{MPC} for modeling the variance.
Specifically, we use the definition of the excess path length describing an ellipse with foci at (virtual) transmitting and receiving nodes, see~\eqref{eq:excess_TX} and~\eqref{eq:excess_RX}.
By introducing a threshold $\xi_{\text{th}}$ for the excess path length, we therefore achieve an inherent geometric separation, i.e., between locations near the propagation path with excess path lengths below the threshold and locations far from the propagation path with excess path lengths above it.
Thus, we model the standard deviation $\sigma_{l,n}(\boldsymbol{r})$ piece-wise as
\begin{equation}
\label{eq:residual_model_binary}
    \sigma_{l,n}(\boldsymbol{r}) =
    \begin{cases} 
      \sigma_{1 \, l,n} \, , & \xi_{l,n}^{\text{min}}(\boldsymbol{r}) > \xi_{\text{th}} \\ \addlinespace
      \sigma_{2 \, l,n} = \sigma_{1 \, l,n} + \Delta \sigma_{l,n} \, , & \xi_{l,n}^{\text{min}}(\boldsymbol{r}) \le \xi_{\text{th}}
   \end{cases}
\end{equation}
where the additive term $\Delta \sigma_{l,n}>0$ reflects the increasing variations in power when a user is near a propagation path and implies that $\sigma_{1 \, l,n} < \sigma_{2 \, l,n}$.
As mentioned above, the excess path length geometrically defines an ellipse with the transceiving nodes as foci.
Similarly, Fresnel zones are defined as a series of discretized concentric ellipses, also with the transceiving nodes as foci~\cite{rappaport1996wireless,zhang2017}.
Therefore, the definition of the Fresnel zones is basically equivalent to that of the excess path length, which allows to define a physically motivated threshold $\xi_{\text{th}}$ as
\begin{equation}
\label{eq:fres_threshold}
  \xi_{\text{th}}  = n_{\text{F}} \, \frac{\lambda}{2} \, , 
\end{equation}
with wavelength $\lambda$ and the number of the Fresnel zone $n_{\text{F}}$.
\begin{table}[t]
    \centering
    \renewcommand{\arraystretch}{1.2} % Default value: 1
    \caption{Standard deviation of noise models for pedestrian user according to \eqref{eq:residual_model_unary} or \eqref{eq:residual_model_binary}}
    \label{tab:model_noise}
    \begin{tabular}{lcccc}
        \toprule
        \multirow{2}{*}{} &  \textbf{Normal \eqref{eq:residual_model_unary}} & \multicolumn{3}{c}{\textbf{Location dependent \eqref{eq:residual_model_binary}}} \\
        \cmidrule(lr){2-2} 
        \cmidrule(lr){3-5} 
%        \cmidrule(lr){3-5} \cmidrule(lr){6-6} \cmidrule(lr){7-7} 
        & $\bar{\sigma}$ [\SI{}{\dB}] & $\xi_{\text{th}}$ [\SI{}{\m}] & $\sigma_1$ [\SI{}{\dB}] & $\sigma_2$ [\SI{}{\dB}] \\ 
        \midrule \addlinespace
        \textbf{Setup I} & \multicolumn{1}{p{1.4cm}}{\centering {0.4659}} & \multicolumn{1}{p{1.2cm}}{\centering {0.0865}} & \multicolumn{1}{p{1.2cm}}{\centering {0.2813}} &  \multicolumn{1}{p{1.2cm}}{\centering {0.7957}} \\  \addlinespace
        \textbf{Setup II} & 1.6630 & 0.0865 & 0.8749 & 2.5683 \\   \addlinespace
        \multicolumn{1}{p{1.4cm}}{{\textbf{Setup III}}} & 2.0252 & 0.1126 & 1.4196 & 2.4534 \\   \addlinespace
        \bottomrule
    \end{tabular}
\end{table}
The corresponding maximum radius of the Fresnel zone is given by 
\begin{equation}
\label{eq:fres_radius}
  r_{l,n}^\text{max}  = \frac{\sqrt{n_{\text{F}}\, \lambda \,d_{l,n}}}{2}.
\end{equation}
On the basis of the measurement data for pedestrian users, we can observe that the standard deviation remains stable for excess path lengths above a threshold corresponding to the third Fresnel zone, i.e., $n_{\text{F}}=3$.
As an example, Fig.~\ref{fig:model_residual} shows the residuals, the threshold, and the course of the standard deviation for Setup~I.
%For Setup~I, Fig.~\ref{fig:model_residual} highlights the separation of the residuals by that threshold. 
While the standard deviation of the residuals corresponding to an excess path length below the threshold increases, the standard deviation of the residuals corresponding to an excess path length above the threshold does not substantially change.
%the standard deviation is stable for residuals of high.
Since the course of the standard deviation is qualitatively comparable for all measurement setups, the threshold calculated with $n_{\text{F}}=3$ also applies to Setup~II and Setup~III.
Note that the value of the threshold for Setup~III differs from that of Setup~I~and~II due to the different center frequencies of the measurement systems (cf.~Table~\ref{tab:meas_overview}).
%Note that this threshold similarly applies to the measurement data of Setup~II and Setup~III.
Finally, Table~\ref{tab:model_noise} gives an overview of the standard deviation values calculated for the pedestrian users from the measurements of Setups~I-III.
As expected, the values for the location dependent noise model clearly indicate that the standard deviation is considerably higher when the user is close to the propagation path.
This is visually confirmed by the two \acp{PDF} of the residuals in Fig.~\ref{fig:model_pdf}, which are obtained for excess path length above and below the threshold ($\mytikzONE$ and $\mytikzTWO$).
%This is visually confirmed in Fig.~\ref{fig:model_pdf}, where the \ac{PDF} of the residuals divided by the threshold are shown individually.
%Comparing the empirical \acp{PDF} with the corresponding fitted normal distributions we can observe a strong fit which further confirms the approximation of the piece-wise defined standard deviation.
The good agreement between the empirical \acp{PDF} with the corresponding fitted normal distributions further confirms the approximation of the noise by the piece-wise defined standard deviation in~\eqref{eq:residual_model_binary}.
%Moreover, comparing the empirical \acp{PDF} with the corresponding fitted normal distributions shown in Fig.~\ref{fig:model_pdf}, we can observe a good agreement, which further confirms the approximation of the piece-wise defined standard deviation in \eqref{eq:residual_model_binary}.

\subsection{Discussion}
\begin{figure}[t]
    \centering
   	\includegraphics[width=3.49in]{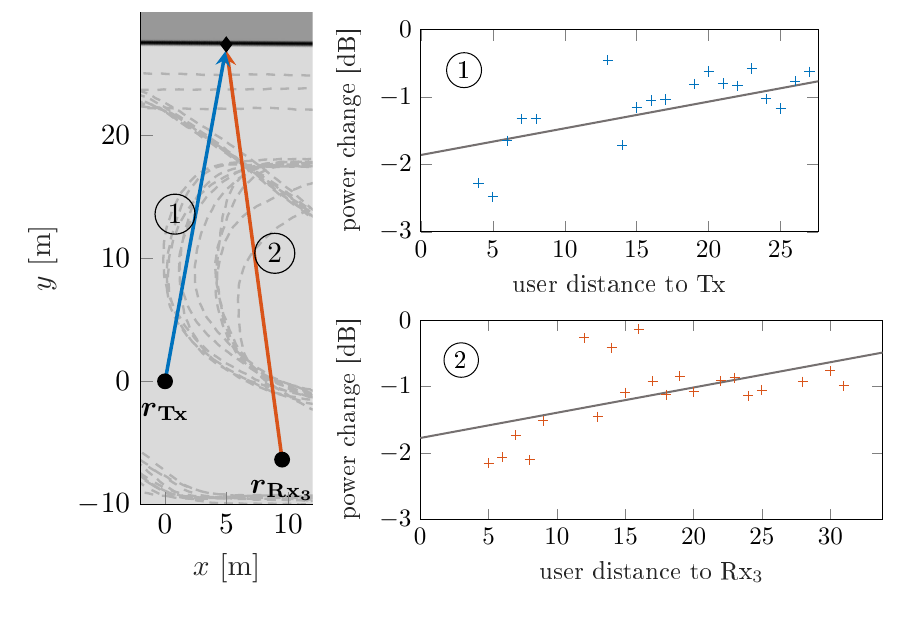}
  	\caption{Exemplary specular reflection for the nodes at $\boldsymbol{r}_{\text{Tx}}$ and $\boldsymbol{r}_{\text{Rx}_3}$ of Setup~I and averaged power changes induced by the pedestrian user when located on the propagation path depending on the user's distance to $\boldsymbol{r}_{\text{Tx}}$~($\mytikzONE$) or~$\boldsymbol{r}_{\text{Rx}_3}$~($\mytikzTWO$).}
  \label{fig:mpc_example} 
\end{figure}
The fitted model parameters for all measured setups are provided in Table~\ref{tab:model_exp} and the corresponding values of the additive noise in Table~\ref{tab:model_noise}.
In the following, we would like to comment on two observations in more detail.

First, when comparing the model parameters for the different user types of Setup~I (cf. Table~\ref{tab:model_exp}), we can see that the values of both the maximum modeled power change $\phi$ and the decay rate $\kappa$ are highest for the car, followed by the bicycle, and lowest for the pedestrian.
Accordingly, the values of the model parameters are related to the dimensions of the user, which is consistent to theoretical models based on diffraction theory, e.g., \cite{wang2015,rampa2015,rappaport2017} for \ac{LoS} paths or \cite{schmidhammer2020} for \acp{MPC}.
Since the dimensions of the car are larger than those of the bicycle, which in turn are larger than those of the pedestrian, we can observe that the absolute induced fading on the received power increases with the dimensions of the user.
The increase in decay rate can be also explained by the user dimensions, as users are modeled empirically as point masses.
Since perturbations in the received power can be measured when the edge of the user approaches the propagation path, the distance of the center point of the user to the propagation path increases with the user's dimensions.
This relation is considered by the model with an increasing decay rate.
%Thus, regarding the center point of the user, the impact on the received power geometrically increases with the dimensions of the user, which is considered by the model with an increasing decay rate.
Hence, the threshold on the excess path length \eqref{eq:fres_threshold} needs to be determined individually for different types of users.
In this work we have determined the threshold on the excess path length for pedestrian users only.
%Considering theoretical models for user induced fading, often based on diffraction theory, e.g.,
%\cite{wang2015,rampa2015,rappaport2017}
%for the \ac{LoS} or \cite{schmidhammer2020} for \acp{MPC}, the following influencing factors can be identified:  wavelength, network geometry and network dimensions, as well as user location and user dimensions.

%
Second, regarding the maximum modeled power change for pedestrian users (cf. Table~\ref{tab:model_exp}), we observe higher fitted values for Setup~II and Setup~III compared to Setup~I.
To explain these values, we refer to the basics of diffraction theory~\cite{rappaport1996wireless}.
The impact on the received power can be explained by the area of the first Fresnel zone blocked by the user~\cite{savazzi2014,rappaport1996wireless}.
Due to the ellipsoidal definition of the Fresnel zone with the maximum radius depending on the length of the propagation path~\eqref{eq:fres_radius}, the intensity of the induced fading depends where the user moves through the propagation path.
As an example, Fig.~\ref{fig:mpc_example} shows the power changes induced by a pedestrian user versus the user's distance to the respective nodes. 
It can be seen, that the user induces a stronger attenuation when being close to the nodes, i.e., when the user relatively blocks a larger part of the Fresnel zones.
This applies both to the path between transmitting node and \ac{RP} and between receiving node and \ac{RP}.
Looking at the user trajectory in Setup~II, cf. Fig.~\ref{fig:meas_techlab}, we see that for the most part the user circled tightly around the network nodes, resulting in comparatively stronger attenuation, which is reflected in the model parameters.
For Setup~III, the smaller dimensions of the network geometry lead to comparatively smaller maximum Fresnel radii~\eqref{eq:fres_radius}.
Therefore, the area blocked by the user is relatively larger resulting in a higher maximum modeled power change.

Referring again to the results in Fig.~\ref{fig:mpc_example}, we can see that the power changes are comparable both qualitatively and quantitatively for both parts of the propagation path. 
%Refering back to Fig.~\ref{fig:mpc_example}, it can be pointed out that the power changes are comparable both qualitatively and quantitatively for both parts of the propagation path. 
%In particular, the course of the measured power changes are both qualitatively and quantitatively comparable for both paths of the propagation path. 
This is an important finding since the user fading is thus independent of which part of the propagation path is affected.
That means, it does not matter whether the user crosses the path between transmitting node and \ac{RP} or between receiving node and \ac{RP}.
This independence of user-induced fading from the respective parts of the propagation path is a prerequisite for the geometric decomposition and thus fundamental to the fading model described in Sec.~\ref{sec:empirical_model}.
Overall, we have therefore introduced a computationally efficient yet accurate exponential fading model for \acp{MPC} that additionally accounts for user location dependent noise through a physically motivated spatial segmentation.

% considering wavelength and network geometry/dimensions
%In the empirical fading model presented above, we account for both the signal specific factor of wavelength and the network dimensions by the threshold on the excess path length defined by the Fresnel zone (cf.~\eqref{eq:fres_threshold}).

% considering wavelength for UWB
%For calculating the 
%With regard to the influence the wavelength of \ac{UWB}
%Regarding the influence of the wavelength for \ac{UWB}
%another point worth pointing out is the 
%\TODO{higher variance due to overlapping Fresnel zones for UWB!}

\section{Bayesian Mapping of Reflection Points}
\label{sec:bayes_mapping}
\begin{figure*}[t]
    \centering
   	\includegraphics[width=6.95in]{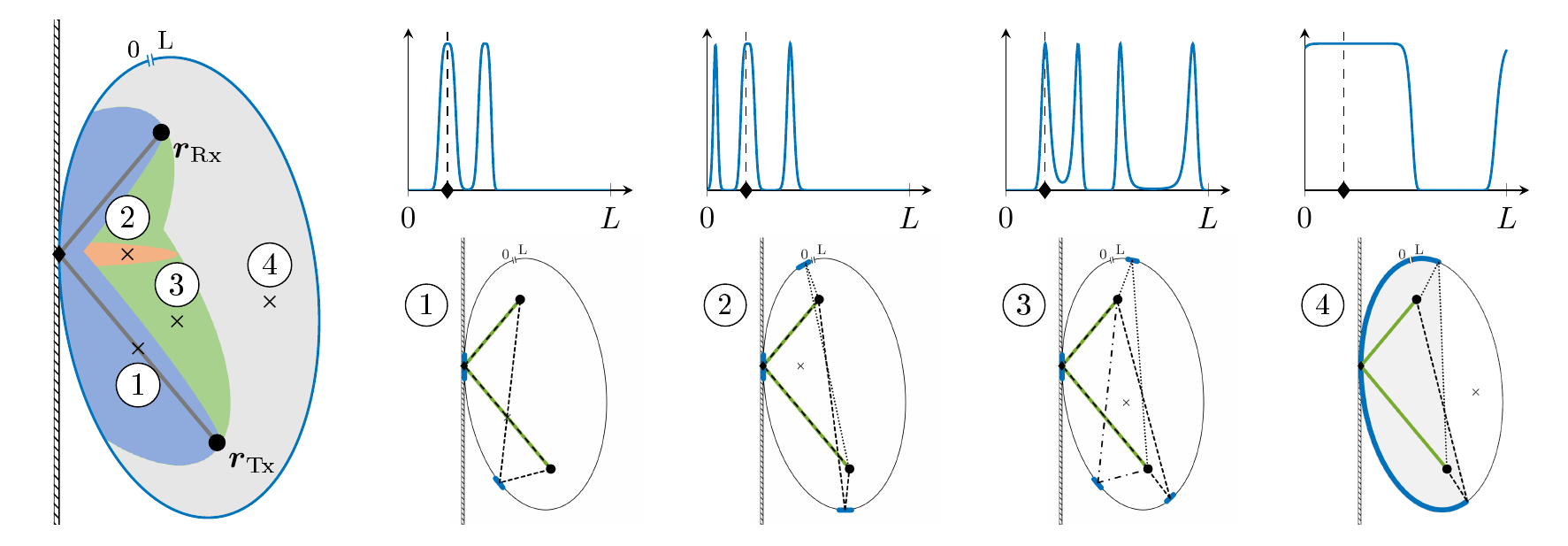}
  	\caption{Exemplary network link and propagation path of an \ac{MPC} due to \acl{SBR}, cf. Fig.~\ref{fig:vt_vr_sbr}. 
  	  Areas of characteristic shapes of the \ac{PDF} are indicated in color.
  	  Corresponding \acp{PDF} on the arc of the delay ellipse are provided for user locations $\mytikzONE$ - $\mytikzFOUR$.
  	  The user locations are indicated by $\times$, the true \ac{RP} is indicated by black diamond, and the \ac{PDF} is highlighted in blue.
  	  }
%  	Colored areas indicate characteristic shapes of the likelihood on the arc of the delay ellipse, which are provided for positions \mytikzONE-\mytikzFOUR.}
  \label{fig:likelihood} 
\end{figure*}
The localization performance of an \ac{MDFL} system is severely depending on the information about the propagation paths within the network.
While the propagation paths for the \ac{LoS} components are given a priori with the locations of the network nodes, the propagation paths of \acp{MPC} still have to be determined.
Therefore, in the following we present a sequential Bayesian estimation approach that determines the location of an \ac{RP} that ultimately defines the propagation path of an \ac{MPC}.

\subsection{Problem Formulation}
\label{sec:mapRP_problem}
For estimating the \acp{RP} we can use the geometric properties of \acp{MPC} caused by \ac{SBR} as described in Sec.~\ref{sec:preGeo}.
That is, the possible locations of an \ac{RP} are described by a delay ellipse, cf.~\eqref{eq:ell_eq_polar}.
Thus, we can uniquely describe the location of the \ac{RP} using the arc length as defined in~\eqref{eq:ell_arc}.
For the $n$-th \ac{MPC} of network link $l$, we can therewith define a \ac{1D} state at time instant $k$ as
\begin{equation}
    \label{eq:RPest_state_user}
    \mathrm{x}_{k} = \left[s_{l,n}(\theta_{k})\right],
\end{equation}
with the arc length $s_{l,n}(\theta_{k})$.
%with the arc length $s_{l,n}(\theta_{k})$ as defined in~\eqref{eq:ell_arc}.
Since the \acp{RP} can be estimated independently, we omit the indices for \ac{MPC} and network link in the following description for notational convenience.
Considering a static \ac{MPC}, we expect the location of the \ac{RP} to be stationary, i.e., the state $\mathrm{x}_{k}$ is time-invariant.
%Thus, with appropriate application of the white noise acceleration model~\cite{bar2004}, the equation of state is
%\begin{equation}
%    \label{eq:RPest_transition_model}
%    \mathrm{x}_{k} = \mathrm{x}_{k-1} + \mathrm{n}_{k},
%\end{equation}
%with zero-mean white Gaussian process noise 
%${\mathrm{n}_{k} \sim \mathcal{N}(0,\sigma^2_{\mathrm{n}})}$.
%The variance of the process noise is defined as 
%$\sigma^2_{\mathrm{n}} = \sigma^2_{\mathrm{p}}\,T_{\mathrm{g}}^3/3$ 
%with the update time $T_{\mathrm{g}}$ between two adjacent measurements and the process noise intensity $\sigma^2_{\mathrm{p}}$ of physical dimension~$\left[\SI[per-mode=symbol]{}{\meter\squared\per\s\tothe{3}}\right]$,
%which is specified according to application requirements~\cite{bar2004}.
The transition prior distribution is thus defined as
\begin{equation}
    \label{eq:RPest_distribution_transition}
    p(\mathrm{x}_{k}|\mathrm{x}_{k-1}) = \delta(\mathrm{x}_{k}-\mathrm{x}_{k-1}),
\end{equation}
where $\delta(\cdot)$ denotes the Dirac delta distribution.

For calibrating, i.e., initially estimating the locations of the \acp{RP}, the \ac{MDFL} system continuously measures the variations in the power of the \ac{MPC}, cf.~\eqref{eq:amplitude_ChangePower}.
Assuming a known user location $\boldsymbol{r}_{k}$, we can model the measured power change equivalently to~\eqref{eq:residual_model_r_dep} as
\begin{equation}
    \label{eq:RPest_meas_model}
    \mathrm{z}_{k} = f(\mathrm{x}_{k},\boldsymbol{r}_{k}) + \mathrm{w}(\boldsymbol{r}_{k}),
    %\mathrm{z}_{k} = f(\mathrm{x}_{k}|\boldsymbol{r}) + \mathrm{w}_{k}(\boldsymbol{r}),
\end{equation}
assuming zero-mean Gaussian distributed measurement noise 
$\mathrm{w}(\boldsymbol{r}_{k}) \sim \mathcal{N}(0,\sigma^2(\boldsymbol{r}_{k}))$
with $\sigma(\boldsymbol{r}_{k})$ as defined in~\eqref{eq:residual_model_binary}.
Applying~\eqref{eq:exp_model_sumMPC}, we can express the power change depending on the location of the \ac{RP} as
\begin{equation}
\label{eq:RPest_exp_model_sumMPC}
f(\mathrm{x}_{k},\boldsymbol{r}_{k}) =  \phi \left(e^{-\xi^{\text{Tx}}(\mathrm{x}_{k},\boldsymbol{r}_{k})/\kappa} + e^{-\xi^{\text{Rx}}(\mathrm{x}_{k},\boldsymbol{r}_{k})/\kappa} \right),
%f(\mathrm{x}_{k}|\boldsymbol{r}) =  \phi \left(e^{-\delta^{\text{Tx}}(\mathrm{x}_{k}|\boldsymbol{r})/\kappa} + e^{-\delta^{\text{Rx}}(\mathrm{x}_{k}|\boldsymbol{r})/\kappa} \right),
\end{equation}
with maximum modeled power change $\phi$ and decay rate $\kappa$.
The user state dependent excess path lengths are defined similarly to~\eqref{eq:excess_TX} and~\eqref{eq:excess_RX} as
\begin{equation}
\label{eq:RPest_excess_TX}
\xi^{\text{Tx}}(\mathrm{x}_{k},\boldsymbol{r}_{k}) =
\Vert \boldsymbol{r}_{\mathrm{Tx}} - \boldsymbol{r}_{k} \Vert + \Vert \boldsymbol{r}_{\mathrm{VR}}(\mathrm{x}_{k}) - \boldsymbol{r}_{k} \Vert - d,
\end{equation}
and
\begin{equation}
\label{eq:RPest_excess_RX}
\xi^{\text{Rx}}(\mathrm{x}_{k},\boldsymbol{r}_{k}) =
\Vert \boldsymbol{r}_{\mathrm{VT}}(\mathrm{x}_{k}) - \boldsymbol{r}_{k} \Vert + \Vert \boldsymbol{r}_{\mathrm{Rx}} - \boldsymbol{r}_{k} \Vert - d,
\end{equation}
where $\boldsymbol{r}_{\mathrm{Tx}}$ and $\boldsymbol{r}_{\mathrm{Rx}}$ denote the locations of the transceiving nodes, respectively.
The geometric length of the propagation path is determined by the propagation delay $\tau$ of the considered \ac{MPC} as in~\eqref{eq:prop_dist}, i.e., $d = c\,\tau$.
The locations of the virtual nodes are calculated according to~\eqref{eq:VT} and~\eqref{eq:VR} as
\begin{equation}
\label{eq:RPest_VT}
\boldsymbol{r}_{\mathrm{VT}}(\mathrm{x}_{k}) = \frac{\boldsymbol{r}_{\mathrm{RP}}(\mathrm{x}_{k})-\boldsymbol{r}_{\mathrm{Rx}}}{\lVert \boldsymbol{r}_{\mathrm{RP}}(\mathrm{x}_{k})-\boldsymbol{r}_{\mathrm{Rx}}\rVert} d + \boldsymbol{r}_{\mathrm{Rx}},
\end{equation}
and
\begin{equation}
\label{eq:RPest_VR}
\boldsymbol{r}_{\mathrm{VR}}(\mathrm{x}_{k}) = \frac{\boldsymbol{r}_{\mathrm{RP}}(\mathrm{x}_{k})-\boldsymbol{r}_{\mathrm{Tx}}}{\lVert \boldsymbol{r}_{\mathrm{RP}}(\mathrm{x}_{k})-\boldsymbol{r}_{\mathrm{Tx}}\rVert} d + \boldsymbol{r}_{\mathrm{Tx}}.
\end{equation}
The location of the \ac{RP} $\boldsymbol{r}_{\mathrm{RP}}(\mathrm{x}_{k})$, expressed in the coordinate system used for the \ac{MDFL} network, is calculated using~\eqref{eq:cart_ell}, i.e., by rotating and translating the location of the \ac{RP} 
$\boldsymbol{r}^{\prime}_{\mathrm{RP}}(\theta(\mathrm{x}_{k}))$, 
% $\boldsymbol{r}^{\prime}_{\mathrm{RP}}(\theta_{k})$, 
expressed in the local ellipse-centric coordinate system.
Finally, we can calculate the ellipse parameter depending on the user state by the inverse function of the arc length defined in~\eqref{eq:ell_arc} as
\begin{equation}
\label{eq:RPest_invers_s}
\theta_{k} = \theta(\mathrm{x}_{k}) = s^{-1}(\mathrm{x}_{k}).
% \theta_{k} = s^{-1}(\mathrm{x}_{k}).
\end{equation}
Note that there is no closed form solution for the inverse function of the arc length, i.e., the ellipse parameter needs to be determined numerically~\cite{fukushima2013}.

\subsection{Elliptic Probability Density Function}
\label{sec:mapRP_pdf}
Due to the \ac{1D} definition of the state $\mathrm{x}_{k}$, i.e., as elliptical arc length, cf.~\eqref{eq:RPest_state_user}, the corresponding \ac{PDF} of state $\mathrm{x}_{k}$ is circularly distributed on the arc of the delay ellipse.
In the following we outline the characteristics of that elliptic \ac{PDF}.
%\TODO{
%The calibration approach presented above aims to estimate the location of an \ac{RP} in terms of the state $\mathrm{x}_{k}$ defined in \ac{1D}, i.e., as an arc length on an ellipse~\eqref{eq:RPest_state_user}.
%For estimation, we therefore need to determine the \ac{PDF} of $\mathrm{x}_{k}$, which is distributed circularly on the arc of the delay ellipse. }
%Due to the \ac{1D} definition of the state $\mathrm{x}_{k}$, i.e., as an arc length on an ellipse,
%the \ac{PDF} of state $\mathrm{x}_{k}$ is distributed circularly on the arc of the delay ellipse.
Therefore, Fig.~\ref{fig:likelihood} illustrates different examples for the elliptical \ac{PDF} using the introductory example of an \ac{MPC} due to \ac{SBR}~(see~Fig.~\ref{fig:vt_vr_sbr}).
As can be seen, the \ac{PDF} can take four different shapes depending on the user's location, which are explained by the measurement model~\eqref{eq:RPest_meas_model} together with the definitions of the excess path lengths in~\eqref{eq:RPest_excess_TX} and~\eqref{eq:RPest_excess_RX}.
Consider the example of a measurement corresponding to an excess path length of $\xi^{\text{Tx}}(\mathrm{x}_{k},\boldsymbol{r}_{k}) = \xi^{\text{Rx}}(\mathrm{x}_{k},\boldsymbol{r}_{k}) = 0$, i.e., the user is located on or very close to the propagation path, as shown in Fig.~\ref{fig:likelihood} for user location $\mytikzONE$ (blue area).
According to the excess path length, we know that the user needs to be located between the \ac{RP} and one of the two transceiving nodes.
As illustrated for user location $\mytikzONE$, there are exactly two possible locations for the \ac{RP}. 
Namely, when the user is located between $\boldsymbol{r}_{\mathrm{Tx}}$ and the location of the true \ac{RP} or, equally likely, when the user is located between $\boldsymbol{r}_{\mathrm{Rx}}$ and a second location of the \ac{RP}.
In the same way, we can explain the three remaining shapes of the \ac{PDF}.
Thereby, for non-zero measurements, the most general case is given for user location $\mytikzTHREE$ (green area).
Here, a measured power change corresponds to a non-zero excess path length, i.e., $\xi^{\text{Tx}}(\mathrm{x}_{k},\boldsymbol{r}_{k}) = \xi^{\text{Rx}}(\mathrm{x}_{k},\boldsymbol{r}_{k}) \neq 0$.
Thus, for each part of the propagation path, i.e. between $\boldsymbol{r}_{\mathrm{Tx}}$ or $\boldsymbol{r}_{\mathrm{Rx}}$ and an \ac{RP}, two individual \ac{RP} locations can be determined, resulting in a total of four possible locations.
This is also true for user location $\mytikzTWO$ (orange area), but due to symmetry, for each part of the propagation path, one of the two determined \ac{RP} locations corresponds to the true \ac{RP}, resulting in three possible locations for the \ac{RP}.
%That means, the three-modal \ac{PDF} is obtained when the user is on the perpendicular to the reflecting surface that passes through the true \ac{RP} .
Finally, a user at location $\mytikzFOUR$ (gray area) does not affect the received power significantly.
Due to the shape of the exponential model, no unique locations for the \ac{RP} can be determined when the measurement is close to zero.
But, depending on the user position, it is possible to distinguish between areas on the elliptic arc in which the \ac{RP} can be located and in which the \ac{RP} can be excluded. 

\subsection{Sequential Bayesian Filtering}
\label{sec:mapRP_bayes}
\begin{algorithm}[t]
    \caption{Bayesian Mapping of \ac{RP} at Time Step~$k$}
    \label{alg:pmf}
    \BlankLine
    \KwIn{\\
        Measured power change: $\mathrm{z}_k$ \\
        User location: $\boldsymbol{r}_k$ \\
        Grid points of PMF: $\{\mathrm{x}^{i}\}_{i=1}^{N_{\mathrm{s}}}$ \\
        Weights of PMF: $\{w_{k-1|k-1}^{i}\}_{i=1}^{N_{\mathrm{s}}}$ \\
    }
    \KwOut{\\
        Weights of PMF: $\{w_{k|k}^{i}\}_{i=1}^{N_{\mathrm{s}}}$ \\
        \acs{MMSE} estimate: $\hat{\mathrm{x}}_k$ \\
    }
    \LinesNumbered
    \If{k=0}
    {
       Initialize weights $w_{k|k}^{i}=1/N_{\mathrm{s}},\quad \forall i\in\{1,\dots,N_{\mathrm{s}}\}$;
    }
    \Else
    {
        \For{i = 1:$N_{\mathrm{s}}$}
        {
            Calculate $w_{k-1|k-1}^{i}$ by circular convolution of $\{w_{k-1|k-1}^{j}\}_{j=1}^{N_{\mathrm{s}}}$ with $p(\mathrm{x}_{}^i|\mathrm{x}_{}^j,\eta)$,~\eqref{eq:RPest_weight_predict}\;
            Calculate $w_{k|k}^{i} = w_{k-1|k-1}^{i} {p}(\mathrm{z}_k|\mathrm{x}_{}^i,\boldsymbol{r}_{k})$,~\eqref{eq:RPest_weight_update}\;
        }
        Normalize weights\;
        % Calculate \ac{MAP} $\hat{\mathrm{x}}_k$ using \eqref{eq:RPest_mmse}\;
        Calculate \acs{MMSE} $\hat{\mathrm{x}}_k$\;
    }
\end{algorithm}
Using a sequential Bayesian estimator, the \ac{PDF} of the state~$\mathrm{x}_{k}$ is determined by computing the posterior density
$p(\mathrm{x}_{k}|\mathrm{z}_{1:k})$
applying the general Bayesian update recursion~\cite{gustafsson2010}.
While for linear system models, such as linear Gaussian systems, the posterior can be efficiently estimated using Kalman filtering solutions, for nonlinear systems the posterior must be numerically approximated in most cases~\cite{siebler2021}.
For the estimation problem outlined in Section~\ref{sec:mapRP_problem}, the nonlinearities are due to the exponential measurement model~\eqref{eq:RPest_meas_model} as well as the state-dependent definition of the excess path lengths~\eqref{eq:RPest_excess_TX} and~\eqref{eq:RPest_excess_RX}. 
Furthermore, due to the properties of the \ac{PDF}, as discussed in~Section~\ref{sec:mapRP_pdf}, the posterior can hardly be assumed to be Gaussian.
Possible filter solutions for such nonlinear and non-Gaussian processes that numerically approximate the posterior are given by the \ac{PF} and the \ac{PMF}~\cite{arulampalam2002,gustafsson2010}.
Given the elliptic \ac{PDF} of the state $\mathrm{x}_{k}$, a \ac{PF} can lead to increased complexity, since according~\cite{kurz2016} sampling from circular distributions can be costly.
In contrast to the \ac{PF}, the \ac{PMF} does not require any resampling.
%, except for an initial sampling step for determining a deterministic grid.
%Except for an initial sampling step for determining a deterministic grid, the \ac{PMF}, in contrast to the \ac{PF}, does not require any resampling.
With respect to complexity, we choose the \ac{PMF} for estimating the posterior.
Therefore, the \ac{PMF} approximates the posterior distribution with the discrete density
\begin{equation}
\label{eq:RPest_pmf}
    {{p}}(\mathrm{x}_{k}|\mathrm{z}_{1:k}) \approx \sum_{i=1}^{N_{\mathrm{s}}} w_{k|k}^i\delta(\mathrm{x}_{k}-\mathrm{x}_{}^i),
\end{equation}
where $\mathrm{x}_{}^i$ represents the $i$-th grid point of the deterministic grid $\{\mathrm{x}_{}^i\}_{i=1}^{N_{\mathrm{s}}}$~\cite{gustafsson2010}.
Due to the definition of the state $\mathrm{x}_{k}$ as arc length, the grid points can be easily arranged equidistantly around the arc of the ellipse. 
The grid point spacing is determined by the circumference of the ellipse $L$, cf.~\eqref{eq:ell_circum_short}, and the number of grid points $N_{\mathrm{s}}$ as
$\Delta_x = L/N_{\mathrm{s}}$.
The weights $w_{k|k}^i$ are calculated as
\begin{equation}
    \label{eq:RPest_weight_update}
    w_{k|k}^i = \frac{1}{c_k}w_{k|k-1}^i {p}(\mathrm{z}_k|\mathrm{x}_{}^i,\boldsymbol{r}_{k}),
\end{equation}
with the normalization term 
$c_k=\sum_{j=1}^{N_{\mathrm{s}}} w_{k|k-1}^j{p}(\mathrm{z}_k|\mathrm{x}_{}^j,\boldsymbol{r}_{k})$ 
and the likelihood distribution ${p}(\mathrm{z}_k|\mathrm{x}_{}^i,\boldsymbol{r}_{k})$, which expresses the measurement model in~\eqref{eq:RPest_meas_model}.
The predicted weights are %given by
\begin{equation}
    \label{eq:RPest_weight_predict}
    w_{k|k-1}^i = \sum_{j=1}^{N_{\mathrm{s}}} w_{k-1|k-1}^j{p}(\mathrm{x}_{}^i|\mathrm{x}_{}^j),
\end{equation}
where ${p}(\mathrm{x}_{}^i|\mathrm{x}_{}^j)$ refers to the transition prior distribution.
Because of the approximation of the posterior density by a discrete density, i.e., by a finite number of stationary grid points, a direct application of the Dirac delta function of~\eqref{eq:RPest_distribution_transition} can lead to estimation problems similar to the problem of loss of diversity~\cite{arulampalam2002}.
Therefore, we approximate the transition prior by an elliptic normal distribution 
% \begin{IEEEeqnarray}{lCl}
% p(\mathrm{x}_{}^i &|&\mathrm{x}_{}^j,\eta)  =  \\
%  &&\frac{a\sqrt{1-\epsilon^2\cos^2\left(\theta(\mathrm{x}_{}^i)\right)}}{L I_0(\eta)}
% \exp\left(\eta\cos\left(2 \pi\frac{\left(\mathrm{x}_{}^i-\mathrm{x}_{}^j\right)}{L}\right)\right)\nonumber\,, 
% \end{IEEEeqnarray}
\begin{equation}
p(\mathrm{x}_{}^i |\mathrm{x}_{}^j,\eta)  = 
 \frac{1}{L I_0(\eta)}
\exp\left(\eta\cos\left(2 \pi\frac{\left(\mathrm{x}_{}^i-\mathrm{x}_{}^j\right)}{L}\right)\right) \,, 
\end{equation}
with~$\eta$ as concentration parameter and $L$ as elliptic circumference~\eqref{eq:ell_circum_short}. $I_0(\cdot)$ is the modified Bessel function of the first kind and order $0$. 
Please refer to the derivation of the elliptic normal distribution in the Appendix.
%Gaussian Kernel $K(\cdot)$ with bandwidth~$\sigma_K$ (see derivation in the Appendix).
%(see derivation in Appendix~\ref{sec:appendixI}).
% Thus,~\eqref{eq:RPest_weight_update} results in
% \begin{equation}
%     \label{eq:RPest_weight_updateK}
%     w_{k|k-1}^i = \sum_{j=1}^{N_{\mathrm{s}}} w_{k-1|k-1}^j{K}(\mathrm{x}_{}^i-\mathrm{x}_{}^j).
% \end{equation}
% Note that applying the Kernel $K(\cdot)$ to the discretized grid points around the elliptic arc, equals a circular convolution~\cite{gustafsson2010,kurz2016}.
Note that applying the elliptic normal distribution to the discretized grid points around the elliptic arc equals a circular convolution~\cite{gustafsson2010,kurz2016}.
% Given the circular posterior density of (\ref{eq:RPest_pmf}), we can finally calculate the \ac{MAP} estimate of the state ${\mathrm{x}}_k$ as
% \begin{equation}
% \label{eq:RPest_mmse}
%     \hat{\mathrm{x}}_k = \max_{\mathrm{x}_{k}} {{p}}(\mathrm{x}_{k}|\mathrm{z}_{1:k}).
% \end{equation}
% \begin{equation}
% \label{eq:RPest_mmse}
%     \hat{\mathrm{x}}_k = \sum_{i=1}^{N_{\mathrm{s}}} w_{k|k}^i \mathrm{x}_{}^i.
% \end{equation}
In order to obtain a point estimate $\hat{\mathrm{x}}_k$ from the elliptic posterior density of~(\ref{eq:RPest_pmf}), we compute the \ac{MMSE} estimate.
Finally, Algorithm~\ref{alg:pmf} provides a pseudocode for the sequential Bayesian filtering approach described above.
%Finally, Algorithm~\ref{alg:pmf} provides a pseudocode for the sequential Bayesian filtering approach described above.
%Lastly, a pseudocode for the sequential Bayesian filtering approach described above can be found in Algorithm~\ref{alg:pmf}.

\section{Experimental Evaluation}
\label{sec:eval}
In this section, we evaluate the above presented Bayesian mapping approach of \acp{RP}.
Therefore, we demonstrate the applicability of the approach using measurement data. 
First, for an exemplary \ac{MPC} for the link between $\text{Tx}$ and $\text{Rx}_1$ of Setup~II.
And second, for estimating all resolvable \acp{MPC} of Setup~I and Setup~III, thus showing the applicability for different environments and different measurement systems.
Prior to the subsequent evaluation, we define the evaluation metric.
For evaluating the \ac{RP} location estimation, we need to consider an error measure taking into account periodicity.
Analogously to the shortest distance on a circle~\cite{kurz2016}, we define the shortest distance on an ellipse as
%calculate the distance error on the arc of the ellipse as
\begin{equation}
    \label{eq:rmseRP}
    \mathrm{e}_{\text{RP}, k} =\min \left( | \hat{\mathrm{x}}_k - \mathrm{x}_{\text{RP}} |, L-| \hat{\mathrm{x}}_k - \mathrm{x}_{\text{RP}} | \right),
%    \mathrm{e}_{\text{RP}, k} = | \hat{\mathrm{x}}_k - \mathrm{x}_{\text{RP}} |,
    %\mathrm{RMSE}_k = \sqrt{ E\left[ \lVert \boldsymbol{r}_{\text{RP}}(\hat{\mathrm{x}}_k) - \boldsymbol{r}_{\text{RP}} \rVert^2\right]},
\end{equation}
where $L$ is the circumference of the ellipse, $\mathrm{x}_{\text{RP}}$ is the arc length of the true \ac{RP}, and $\hat{\mathrm{x}}_k$ is the point estimate of the arc length from the \ac{PMF}.

\begin{figure}[t]
    \centering
   	\includegraphics[width=3.49in]{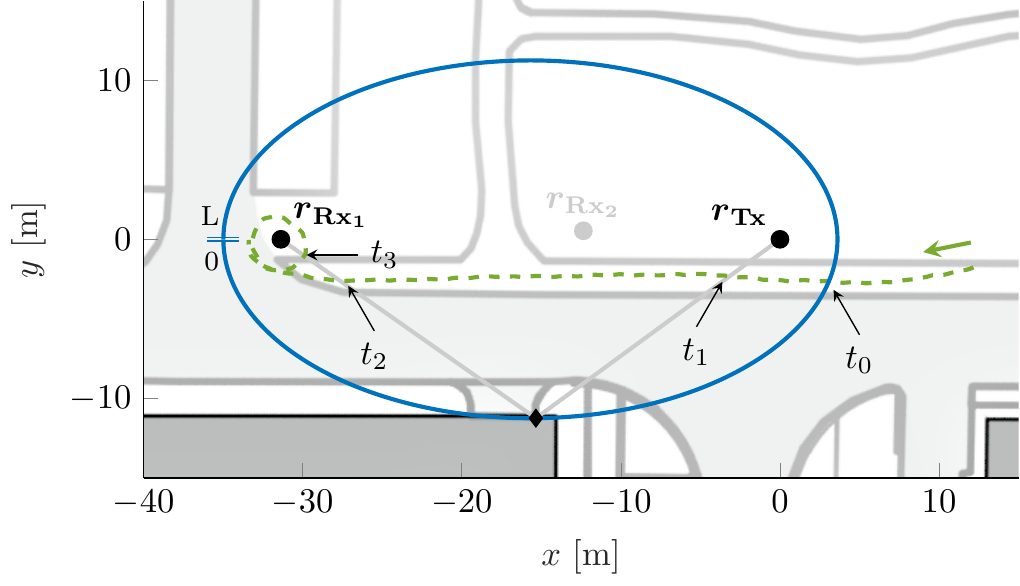}
  	\caption{Overview of measurement environment. Grid points located on delay ellipse indicated in blue, true \ac{RP} indicated by black diamond marker, corresponding propagation path indicated in gray, and trajectory including motion direction highlighted in green.}
  \label{fig:rp_pmf_meas_env} 
\end{figure}
\begin{figure}
    \centering
   	\includegraphics[width=3.49in]{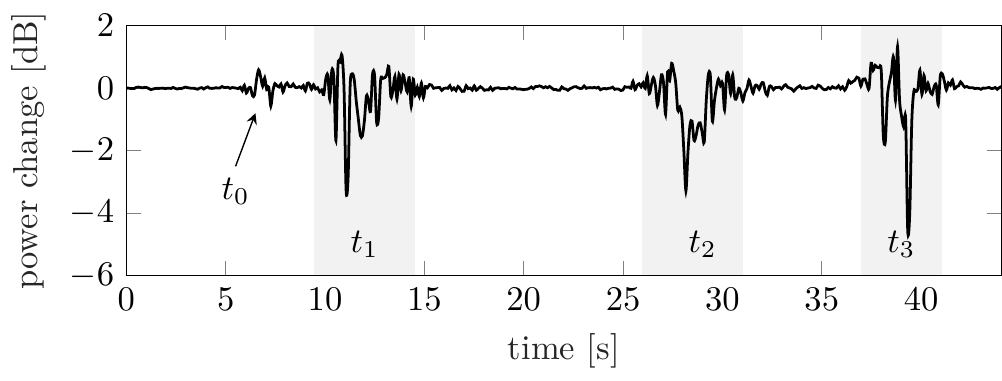}
  	\caption{Measured power changes in \SI{}{\dB} for an \ac{MPC} with propagation distance of \SI{38.673}{\m}. Noticeable variations are highlighted in gray and the corresponding times are labeled as $t_i$,$i\in\{0,1,2,3\}$.}
  \label{fig:rp_pmf_meas_time} 
\end{figure}

%\label{sec:eval_measurements}
First, we demonstrate the Bayesian \ac{RP} mapping approach with measurement data from Setup~II.
Details on the measurement data are given in Table~\ref{tab:meas_overview}.
Specifically, we consider the link between $\text{Tx}$ and $\text{Rx}_1$ with a distance of $d_{\text{LoS}}=\SI{31.370}{\m}$ 
and an \ac{MPC} with an estimated propagation distance of $d = \SI{38.673}{\m}$.
%and an \ac{MPC} with an estimated propagation delay of $\tau = \SI{0.129}{\micro\s}$, i.e., using~\eqref{eq:prop_dist} a distance of $d = \SI{38.673}{\m}$.
As shown in Section~\ref{sec:preGeo}, possible locations for the \ac{RP} are determined by an ellipse, cf.~\eqref{eq:ell_eq_polar}, with principle axis of 
% as given in~\eqref{eq:ell_eq_polar}
% a = d/2
$a = \SI{19.337}{\m}$ and 
% b = sqrt(d^2-d_LoS^2)/2
$b = \SI{11.308}{\m}$, see~\eqref{eq:ell_a} and~\eqref{eq:ell_b}, respectively.
Using~\eqref{eq:ell_circum_short}, the circumference is calculated as $L=\SI{97.633}{\m}$.
An overview of the considered measurement environment is provided in Fig.~\ref{fig:rp_pmf_meas_env}, including the transmitting and receiving nodes and the aforementioned delay ellipse. 
The true \ac{RP} and the propagation path are shown for illustration only.
Furthermore, Fig.~\ref{fig:rp_pmf_meas_env} shows the calibration trajectory of a pedestrian user, here determined using \ac{GNSS} (cf. Table~\ref{tab:meas_overview}).
Note that during calibration the user is assumed to provide the position information to the \ac{MDFL} system.
The power changes of the \ac{MPC} measured while the user was moving are shown in Fig.~\ref{fig:rp_pmf_meas_time}.
Thereby, we can clearly observe three time intervals, labeled as $t_1$-$t_3$, of increased variations of the received power.
The user positions that correspond to these intervals are indicated by arrows in Fig.~\ref{fig:rp_pmf_meas_env}.
It can be clearly seen, that the time intervals of increased variations in the power coincide with the times when the user passes through the propagation path. 
Further, we have pointed to a time instant $t_0$ in both Fig.~\ref{fig:rp_pmf_meas_env} and Fig.~\ref{fig:rp_pmf_meas_time}.
As shown, the user is located on the delay ellipse at $t_0$.
That means that at this user location, the delay of a directly scattered signal coincides with that of the considered \ac{MPC}, which explains the variations in the received power around $t_0$.
These variations are independent of the \ac{RP} that we want to estimate, and thus would certainly affect the performance of the \ac{RP} estimation.
Since we know the user location during calibration, we can easily mitigate such potentially deceptive measurements when the user is near the delay ellipse.

For estimating the \ac{RP}, we realize the \ac{PMF} as introduced in Section~\ref{sec:mapRP_bayes}.
The grid points of the \ac{PMF} are uniformly distributed on the elliptical arc with a spacing of $\Delta_x = \SI{5}{\cm}$, i.e., a total of $N_{\mathrm{s}} = 1953$ grid points.
Initially, the weights of the \ac{PMF} are set equal.
For the measurement model~\eqref{eq:RPest_meas_model}, we choose the parameters
$\phi = \SI{-2.5}{\dB}$ and $\kappa = \SI{0.015}{\dB}$. 
For the weight prediction~\eqref{eq:RPest_weight_predict}, we use a concentration parameter of 
$\eta = 1/\SI[per-mode=symbol]{0.01}{{\mm\tothe{2}}}$ 
%$1/\eta = \SI[per-mode=symbol]{0.01}{{\mm\tothe{2}}}$ 
for the elliptic normal distribution.
As measurement noise, we consider both the normal distributed~\eqref{eq:residual_model_unary}, with $\bar{\sigma} = \SI{0.75}{\dB}$, as a reference, and the user location dependent normal distributed~\eqref{eq:residual_model_binary}, with $\sigma_1 = \SI{0.5}{\dB}$ and $\sigma_2 = \SI{1.0}{\dB}$.

%\begin{table}[t]
%    \centering
%    \renewcommand{\arraystretch}{1.2} % Default value: 1
%    \caption{Measurement geometry and filter parameters}
%    \label{tab:exp_Bayesian_RP}
%    \begin{tabular}{lcc}
%        \toprule
%        \multicolumn{2}{c}{\textbf{Parameter}} & \multicolumn{1}{p{1.8cm}}{\centering \textbf{Value}}  \\
%        \midrule \addlinespace
%        \ac{LoS} propagation distance       &  $d_{\text{LoS}}$ & \SI{31.370}{\m} \\ \addlinespace
%        \ac{MPC} propagation distance       &  $d$              & \SI{38.673}{\m} \\ \addlinespace
%         Grid spacing       &  $\Delta_{x}$              & \SI{5}{\cm} \\ \addlinespace
%         Grid points        &  $N_{\mathrm{s}}$              & 1953 \\ \addlinespace
%        \bottomrule
%    \end{tabular}
%\end{table}

\begin{figure}
    \centering
   	\includegraphics[width=3.49in]{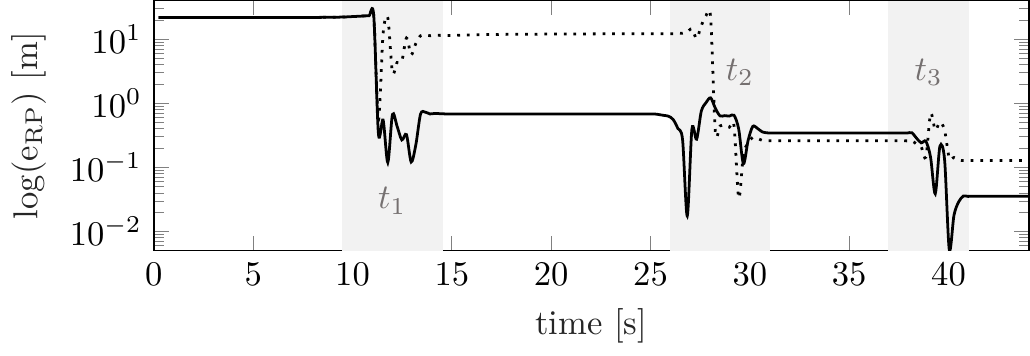}
  	\caption{Distance error of estimated \ac{RP} location over time, for \ac{PMF} realization with noise model of~\eqref{eq:residual_model_unary},~(\mytikzLineDotted), and with user location dependent noise model of~\eqref{eq:residual_model_binary},~(\mytikzLineSolid).
  	  The distance error is given in logarithmic domain.
  	%Vertical dashed lines indicate time instances for PDFs given in Fig.\TODO{!?}.
  	}
  \label{fig:rp_pmf_meas_err} 
\end{figure}

The resulting distance error over time is shown in Fig.~\ref{fig:rp_pmf_meas_err}.
Initially, the weights of the \ac{PMF} are set equal, i.e., the posterior \ac{PDF} corresponds to a uniform distribution.
That uniform distribution in turn explains the initially large distance error.
%The initial high error can be explained by the equally initialized weights,
%since, using~\eqref{eq:RPest_mmse}, the state estimate results in $\hat{\mathrm{x}}_k =L/2 = 48.82$.
%Thus, for the arc length of the true \ac{RP}, i.e., $\mathrm{x}_{\text{RP}} = \SI{24.23}{\m}$, the initial distance error results in $\mathrm{e}_{\text{RP}, 0} = \SI{24.59}{\m}$.
When the user passes the propagation path for the first time ($t_1$), we can observe that the distance error immediately decreases from the initial value.
Here, the \ac{PMF} realized with the location dependent noise model achieves a higher performance gain when the user passes the propagation path the first time. 
When the user passes the propagation path the second time ($t_2$), also the \ac{PMF} realization using the noise model of~\eqref{eq:residual_model_unary} converges.
The results of both \ac{PMF} realization achieve a similar distance error. 
After circling around the receive antenna ($t_3$), the distance errors of both realizations further decrease.
With a final distance error of $\mathrm{e}_{\text{RP}, \text{end}} = \SI{12.8}{\cm}$ using the noise model of~\eqref{eq:residual_model_unary}, and of $\mathrm{e}_{\text{RP}, \text{end}} = \SI{3.5}{\cm}$ using the user location dependent noise model of~\eqref{eq:residual_model_binary}, both realizations achieve an accurate estimation result.

\begin{figure*}[t!]
	\centering
    \hfill
   	\subfloat[Setup~I: controlled apron environment\label{fig:meas_hangar_RP}]{%
	   	\includegraphics[width=3.962in]{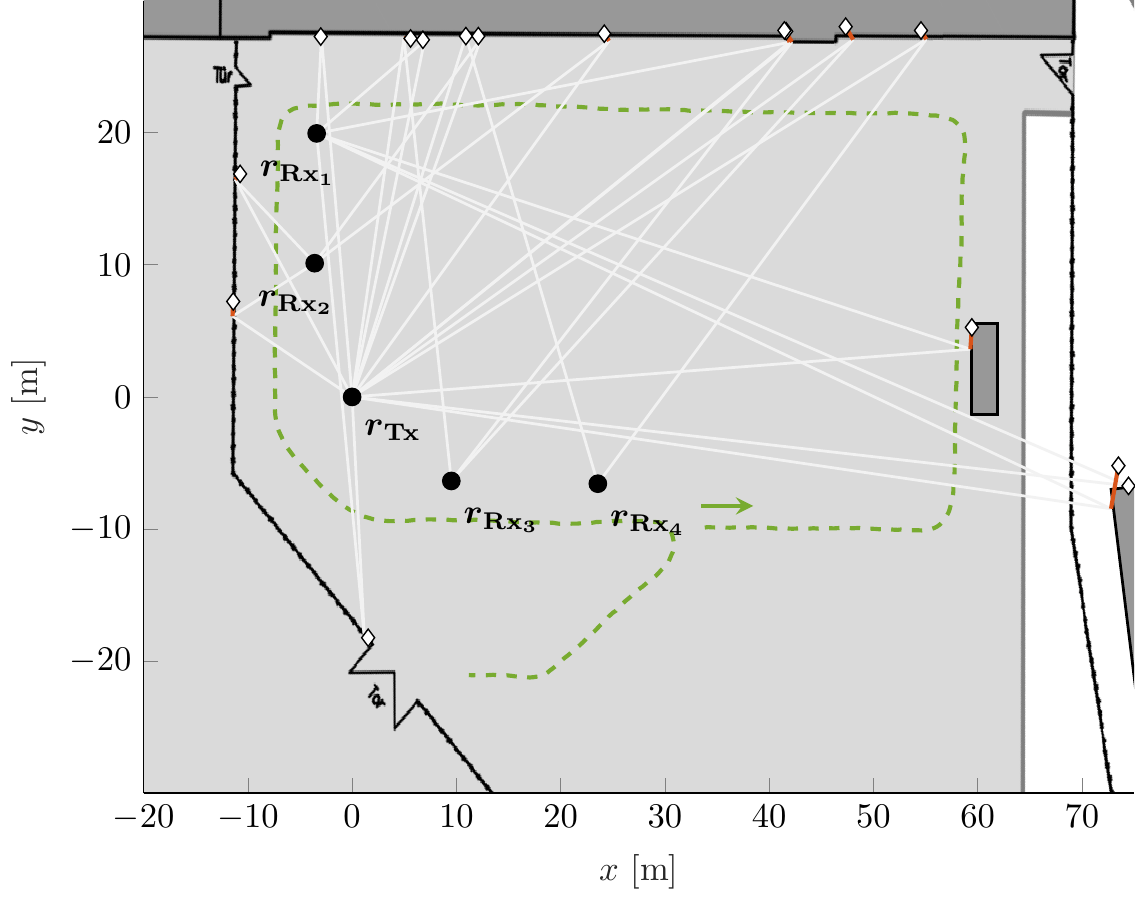}}
    \hfill
	\centering
   	\subfloat[Setup~III: indoor environment\label{fig:meas_holodeck_RP}]{%
	   	\includegraphics[width=2.508in]{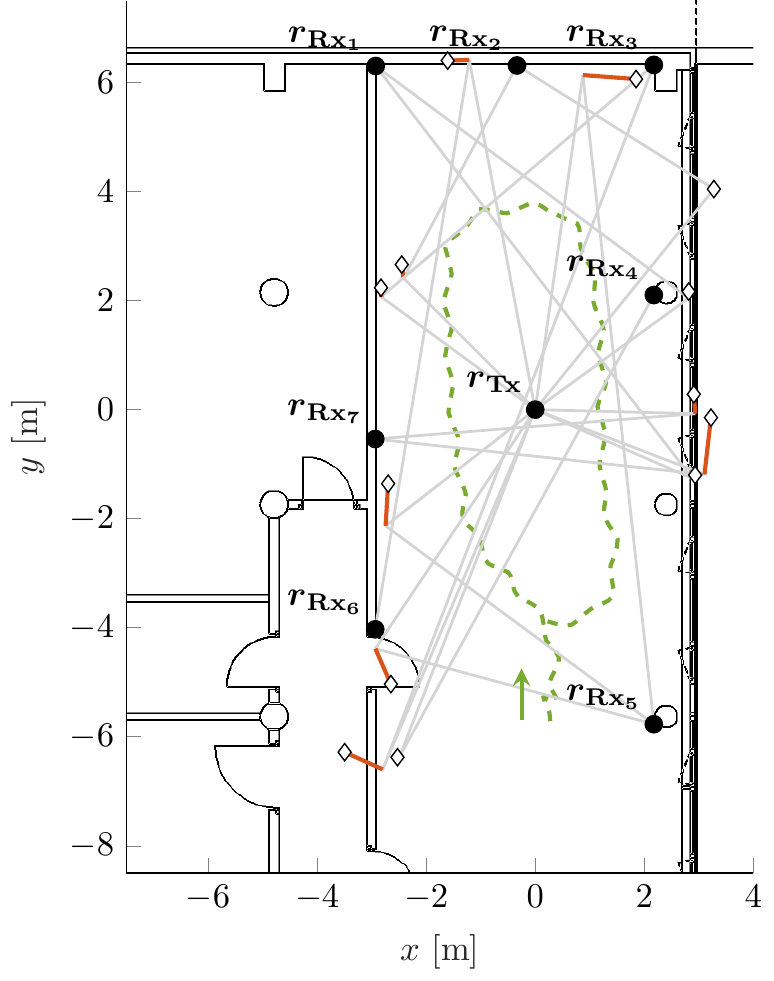}}
    \hfill
	\setlength{\belowcaptionskip}{-5pt}
  	\caption{Resulting \ac{RP} location estimates: true propagation paths are shown as gray lines, estimated \ac{RP} locations are indicated by diamonds, and calibration trajectory is represented by dashed green line for each setup; distance between estimated and true \acp{RP} are highlighted by red lines; averaged distance error of \ac{RP} estimates for (a) Setup~I: $\bar{\mathrm{e}}_{\text{RP}} = \SI{0.82}{\m}$ and for (b) Setup~III: $\bar{\mathrm{e}}_{\text{RP}} = \SI{0.44}{\m}$.
  }
  \label{fig:meas_RP} 
\end{figure*}
In order to further demonstrate the applicability of the Bayesian \ac{RP} mapping approach to measurement data in different environments and for different measurement systems, we apply the approach to the resolvable \acp{MPC} of Setup~I and of Setup~III.
Therefore, we use the same \ac{PMF} realization as introduced above.
Only differences are the number of grid points, that are adjusted individually for each \ac{MPC}, due to different ellipse circumferences and a fixed grid point spacing. 
And second, the concentration parameter of the elliptic normal distribution is adapted for the \ac{UWB} measurement system.
For \ac{UWB}, we set 
$\eta = 1/\SI[per-mode=symbol]{3.24}{{\cm\tothe{2}}}$ 
% $1/\eta = \SI[per-mode=symbol]{3.24}{{\cm\tothe{2}}}$ 
accounting for the increased update time $T_\text{g}$, see Table~\ref{tab:meas_overview}.
Further, we choose the normal distributed noise model of~\eqref{eq:residual_model_unary} for simplicity.
The calibration trajectories for both setups are shown in Fig.~\ref{fig:meas_hangar_RP} and Fig.~\ref{fig:meas_holodeck_RP}, respectively, along with the resulting \ac{RP} location estimates.
%Finally, the resulting \ac{RP} location estimates are illustrated in Fig.~\ref{fig:meas_RP}.
The distance error for each \ac{RP} is visualized in red. 
Overall, the estimated \ac{RP} locations agree very well with the true locations in both setups. 
In particular, we can not identify any substantially incorrect RP location estimates.
The averaged distance errors over all considered \acp{MPC} of the respective setups, i.e., 
$\bar{\mathrm{e}}_{\text{RP}} = \SI{0.82}{\m}$
for Setup~I and 
$\bar{\mathrm{e}}_{\text{RP}} = \SI{0.44}{\m}$
for Setup~III, support the visually observed robust estimation performance.
Note that the higher distance error of Setup~I can be explained by the larger dimensions of the environment of Setup~I compared to Setup~III, which result in larger absolute errors.
In summary, the Bayesian \ac{RP} mapping approach was thus successfully applied to different measurement systems and different environments.

\section{Conclusion}
\label{sec:conclusion}
\acresetall
This paper provides a novel Bayesian calibration approach for \ac{MDFL} systems that determines the propagation paths of \aclp{MPC}.
%This paper provides a novel Bayesian estimation approach for determining the propagation paths of \acl{MPC}s, which automates the calibration of \ac{MDFL} systems.
%Bayesian calibration approach of \ac{MDFL} systems for determining the propagation paths of \acl{MPC}s.
% device-free localization systems that incorporate power measurements from \acl{MPC}s, rely on the information of the propagation paths.
% In this paper, we provide first a statistical fading model that relates changes in the received power of a \acl{MPC} to the location of a user.
Therefore, we first present a statistical fading model that describes user induced changes in the received power of a \acl{MPC}.
% The model is essential for \ac{MDFL} systems, which aim to locate users based on changes in the received power of \acl{MPC}s in addition to these of \acl{LoS} signal components.
% Thus, besides for calibration, the model is also an essential building block for \ac{MDFL} systems 
% that infer the location of a user based on the induced changes in the received power of both \acl{MPC}s and \acl{LoS} signal components.
The model is thus not only important for calibration, but also an essential building block for \ac{MDFL} systems that infer the location of a user based on the induced changes in the received power of both \acl{MPC}s and \acl{LoS} signal components.
%that locate users based on changes in the received power of \acl{MPC}s in addition to these of \acl{LoS} signal components.
Based on an extensive set of wideband and \ac{UWB} measurement data for both indoor and outdoor environments, we derive and validate the model empirically.
%The model is derived and validated based on an extensive set of wideband and \acl{UWB} measurement data for both indoor and outdoor environments.
%An analysis of the measurement data has shown that the presence of users not only attenuates the received power, but also leads to increasing fluctuations.
As emerges from the measurement data, the presence of users not only attenuates the received power, but also leads to increasing fluctuations.
Thus, in addition to the empirical exponential fading model, we propose a location dependent variance model for the measurement noise using an efficient, physically motivated spatial segmentation.

Second, we present the novel Bayesian calibration approach for \ac{MDFL} that robustly estimates the locations of \acl{RP}s from \acl{SBR}s.
%Moreover, we present a novel Bayesian calibration approach for \ac{MDFL} that robustly estimates the locations of \acl{RP}s from \acl{SBR}s.
\ac{MDFL} systems severely depend on the information about the propagation paths within the network.
While known for the \acl{LoS}, the propagation paths have yet to be determined for \acl{MPC}s, which is ultimately achieved by determining corresponding reflection points.
Using the initially derived empirical fading model and given user location during calibration, we can relate measured changes in the received power of the \acl{MPC} to the location of a \acl{RP}.
Taking advantage of geometrical properties of \acl{MPC}s caused by \acl{SBR}s, we can constrain the possible locations of \acl{RP}s to the locations of the corresponding delay ellipse, which allows to formulate a one-dimensional elliptic estimation problem.
We propose the \acl{PMF} that approximates the elliptic posterior density by a deterministic grid for efficiently solving the estimation problem.
The applicability of the presented approach is demonstrated and evaluated using measurement data for different environments and different measurement systems, including a commercial off-the-shelf \ac{UWB} system.
The Bayesian calibration approach is shown to robustly estimate the locations of the \acl{RP}s.
Independent of the considered environment or measurement system, the estimation approach achieves a sub-meter accuracy.

\appendix[Derivation of Elliptic Normal Distribution]
\label{sec:appendixI}
The von Mises distribution, also denoted as circular normal distribution, describes a probability distribution around a circle. 
%It was originally foreseen as atomic weight distribution. 
Its equation is given by \cite{vonmises18}
\begin{equation}
\label{eq:appendix_VM}
p(\alpha|\mu,\kappa)=\frac{1}{2\pi I_0(\kappa)}\exp\left(\kappa\cos\left(\alpha-\mu\right)\right)\,,\quad \alpha\in[0,2\pi)\,,
\end{equation}
with concentration parameter $\kappa$, location parameter $\mu$, and modified Bessel function $I_0(\cdot)$ of the first kind and order $0$.
The von Mises \ac{PDF} is defined on a unit circle.
%Changing the von Mises distribution to arbitrary length circles with circumference $L$, we obtain
We want to transform the \acs{PDF} of $\alpha$ to circles of arbitrary length with circumference $L_{\mathrm{c}}$.
Therefore, we define $s_{\mathrm{c}}$ as the length on the circle corresponding to the angular parameter $\alpha$, with the variable substitution
\begin{equation}
\label{eq:appendix_substi}
\alpha =2\pi\frac{s_{\mathrm{c}}}{L_{\mathrm{c}}}\,.
\end{equation}
The Jacobi element, which is needed for the correct normalization factor according to \cite{Papoulis02} is given by
\begin{equation}
\label{eq:appendix_jacobi}
\frac{\mathrm{d}\alpha}{\mathrm{d}s_{\mathrm{c}}}=\frac{2\pi}{L_{\mathrm{c}}}\,.
\end{equation}
The \ac{PDF} of $s_{\mathrm{c}}$ is obtained by applying the probability density transformation
\begin{equation}
\label{eq:appendix_trafo}
    p(s_{\mathrm{c}}|\bar{s}_{\mathrm{c}},\kappa) = p(\alpha|\mu,\kappa)\left|\frac{\mathrm{d}\alpha}{\mathrm{d}s_{\mathrm{c}}}\right|,
\end{equation}
thus, by inserting~\eqref{eq:appendix_VM} and \eqref{eq:appendix_jacobi} into \eqref{eq:appendix_trafo}, we obtain
\begin{equation}
p(s_{\mathrm{c}}|\bar{s}_{\mathrm{c}},\kappa)=\frac{1}{L_{\mathrm{c}} I_0(\kappa)}\exp\left(\kappa\cos\left(2\pi \frac{\left(s_{\mathrm{c}}-\bar{s}_{\mathrm{c}}\right)}{L_{\mathrm{c}}}\right)\right).
\end{equation}

% Transferring the von Mises distribution to circles of arbitrary length with circumference $L_{\mathrm{c}}$, we obtain
% \begin{equation}
% p(s_{\mathrm{c}}|\bar{s}_{\mathrm{c}},\kappa)=\frac{1}{L_{\mathrm{c}} I_0(\kappa)}\exp\left(\kappa\cos\left(2\pi \frac{\left(s_{\mathrm{c}}-\bar{s}_{\mathrm{c}}\right)}{L_{\mathrm{c}}}\right)\right)
% \end{equation}
% with the variable substitution
% \begin{equation}
% \alpha =2\pi\frac{s_{\mathrm{c}}}{L_{\mathrm{c}}}\,,
% \end{equation}
% where $s_{\mathrm{c}}$ is the length on the circle corresponding to the angular parameter $\alpha$.
% The Jacobi element, which is needed for the correct normalization factor according to \cite{Papoulis02} is given by
% \begin{equation}
% \frac{\mathrm{d}\alpha}{\mathrm{d}s_{\mathrm{c}}}=\frac{2\pi}{L_{\mathrm{c}}}\,.
% \end{equation}
In the following, we want to extend the distribution to an ellipse with circumference $L_{\mathrm{e}}$, cf.~\eqref{eq:ell_circum_short}. 
%with length of an arbitrary point on the ellipse $s$ and the the total length of the ellipse $L$, we obtain
% \begin{equation}
% s_{\mathrm{e}}=a\int_0^{\varphi}\sqrt{1-\epsilon^2\cos^2\zeta}\,\mathrm{d}\zeta 
% \end{equation}
% \begin{equation}
% L_{\mathrm{e}}=4a\int_0^{\pi/2}\sqrt{1-\epsilon^2\cos^2\zeta}\,\mathrm{d}\zeta  =4aE(\epsilon)  
% \end{equation}
With the definition of an arbitrary arc length of the ellipse of
\begin{equation}
\label{eq:appendix_ell_s}
s_{\mathrm{e}} = 
a \int_0^{\theta}\sqrt{1-\epsilon^2 \, \cos^2\vartheta}\,\text{d}\vartheta \,,
\end{equation}
we can define, similarly to \eqref{eq:appendix_substi}, the variable substitution
\begin{equation}
\label{eq:appendix_ell_substi}
\alpha =2\pi\frac{s_{\mathrm{e}}}{L_{\mathrm{e}}}\,.
\end{equation}
Equivalently to \eqref{eq:appendix_jacobi}, the Jacobi element is given by
\begin{equation}
\label{eq:appendix_ell_jacobi}
\frac{\mathrm{d}\alpha}{\mathrm{d}s_{\mathrm{e}}}=\frac{2\pi}{L_{\mathrm{e}}}\,.
\end{equation}
% where differential length for an arbitrary arc length $s_{\mathrm{e}}$ is 
% \begin{equation}
% \mathrm{d}s_{\mathrm{e}}=    
% a\sqrt{1-\epsilon^2\cos^2\vartheta}\,\mathrm{d}\vartheta\,.
% \end{equation}
Similarly to \eqref{eq:appendix_trafo}, the \ac{PDF} of $s_{\mathrm{e}}$ is obtained by applying the probability density transformation
\begin{equation}
\label{eq:appendix_ell_trafo}
    p(s_{\mathrm{e}}|\bar{s}_{\mathrm{e}},\kappa) = p(\alpha|\mu,\kappa)\left|\frac{\mathrm{d}\alpha}{\mathrm{d}s_{\mathrm{e}}}\right|,
\end{equation}
thus, by inserting~\eqref{eq:appendix_VM} and \eqref{eq:appendix_ell_jacobi} into \eqref{eq:appendix_ell_trafo}, we finally obtain the elliptic normal distribution
\begin{equation}
p(s_{\mathrm{e}}|\bar{s}_{\mathrm{e}},\kappa)=\frac{1}{L_{\mathrm{e}} I_0(\kappa)}\exp\left(\kappa\cos\left(2\pi \frac{\left(s_{\mathrm{e}}-\bar{s}_{\mathrm{e}}\right)}{L_{\mathrm{e}}}\right)\right).
\end{equation}

\section*{Acknowledgment}
The presented work was in parts carried out in the \acs{VIDETEC} project funded by the \acl{BMVI} through the \acs{mFUND} initiative under the grant 19F1074A. 
The authors would like to thank Siwei Zhang, Armin Dammann, and Christoph Schmidhammer for fruitful discussions on statistical modelling and Bayesian estimation.

\balance
% references section
\bibliographystyle{IEEEtran}
\bibliography{bib_v1}

\end{document}